%% file: transaction_head.tex
\documentclass[lettersize,journal]{IEEEtran}

\input{settings/packages}

\input{settings/commands}

\usepackage{hologo}
\usepackage{listings}
\begin{document}
	
\pagestyle{empty}
	
\twocolumn[
\begin{@twocolumnfalse}
	\Huge {IEEE copyright notice} \\ \\
	\large {\copyright\ 2024 IEEE. Personal use of this material is permitted. Permission from IEEE must be obtained for all other uses, in any current or future media, including reprinting/republishing this material for advertising or promotional purposes, creating new collective works, for resale or redistribution to servers or lists, or reuse of any copyrighted component of this work in other works.} \\ \\
	
	{\Large Published in \emph{IEEE Transactions on Intelligent Vehicles}, 29 July 2024.} \\ \\
	
	Cite as:
	
	\vspace{0.1cm}
	\noindent\fbox{%
		\parbox{\textwidth}{%
			L.~Ullrich, M.~Buchholz, K.~Dietmayer, and K.~Graichen, ''Expanding the Classical V-Model for the Development of Complex Systems Incorporating AI,''
			in \emph{IEEE Transactions on Intelligent Vehicles}, 29 July 2024, pp. 1--15, doi: 10.1109/TIV.2024.3434515.
		}%
	}
	\vspace{2cm}
	
\end{@twocolumnfalse}
]

\noindent\begin{minipage}{\textwidth}
	
\hologo{BibTeX}:
\footnotesize
\begin{lstlisting}[frame=single]
@article{ullrich2024expanding,
	title={Expanding the Classical V-Model for the Development of Complex Systems Incorporating AI},
	author={Ullrich, Lars and Buchholz, Michael and Dietmayer, Klaus and Graichen, Knut},
	journal={IEEE Transactions on Intelligent Vehicles},
	year={2024},
	pages={1--15},
	doi={10.1109/TIV.2024.3434515},
	publisher={IEEE}
}
\end{lstlisting}
\end{minipage}

\maketitle
\setcounter{page}{1}

\input{src/00_Abstract}
\input{src/01_Introduction}

\input{src/02_StateOfTheArt}

\input{src/03_Methodology}

\input{src/04_Examples}

\input{src/05_Conclusion}

\bibliographystyle{IEEEtran}
\bibliography{literature}
\vspace{-15 mm}
\begin{IEEEbiography}[{\includegraphics[width=1in,height=1.25in,clip,keepaspectratio]{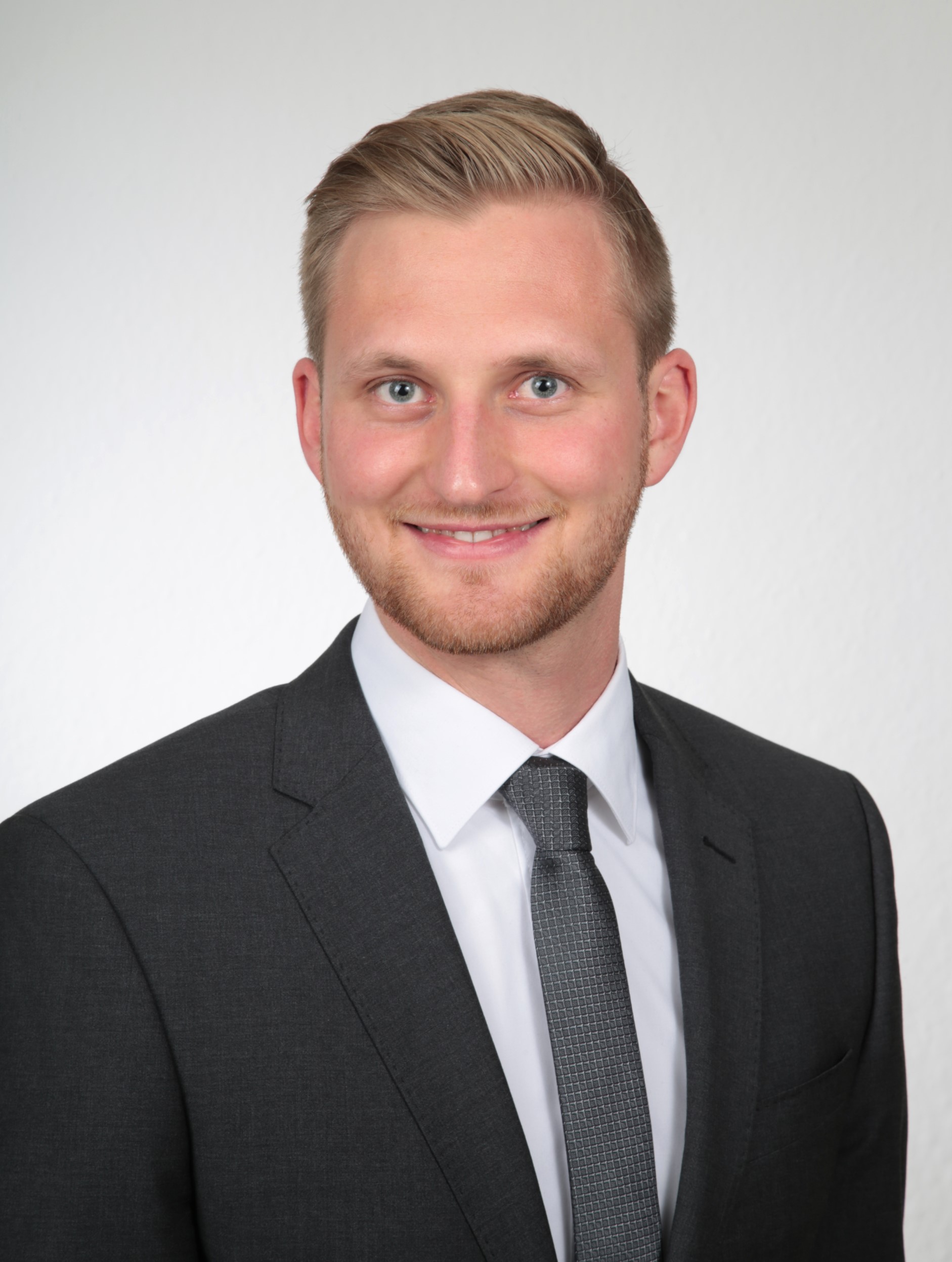}}]{Lars Ullrich}
	received the M.Sc. degree in mechatronics from Friedrich–Alexander–Universita\"at Erlangen–N\"urnberg, Germany, in 2022, where he is currently pursuing the Ph.D.
	(Dr.Ing.) degree with the Chair of Automatic Control.
	
	His research interests include probabilistic trajectory planning for safe and reliable autonomous driving in uncertain dynamic environments with a focus on addressing challenges arising from the use of AI systems in automated driving.
\end{IEEEbiography} \vspace{-15 mm}
\begin{IEEEbiography}[{\includegraphics[width=1in,height=1.25in,clip,keepaspectratio]{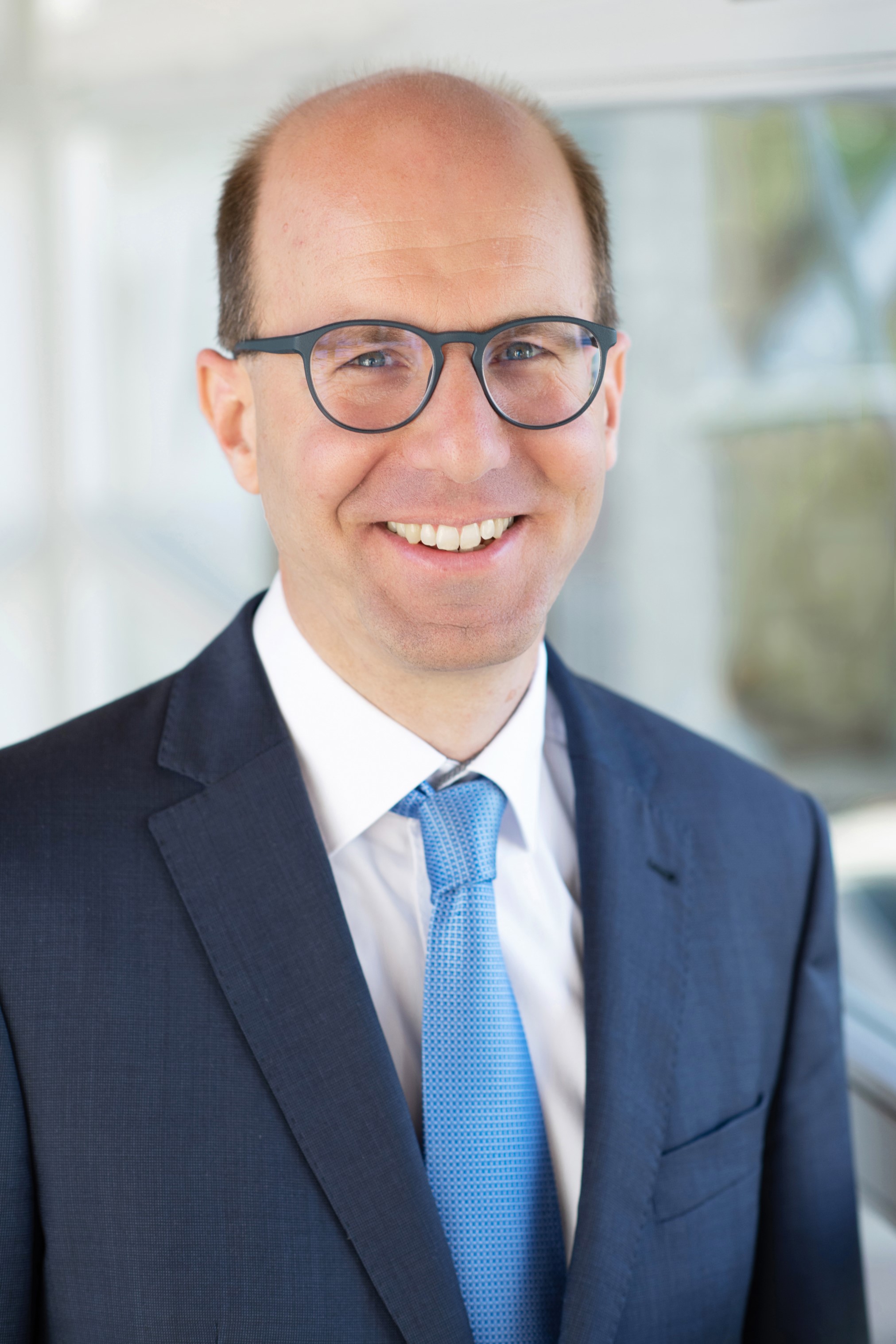}}]{Michael Buchholz}
	received his Diploma degree in Electrical Engineering and Information Technology as well as his Ph.D. from the faculty of Electrical Engineering and Information Technology at University of Karlsruhe (TH)/Karlsruhe Institute of Technology, Germany.  He is a research group leader and lecturer at the Institute of Measurement, Control, and Microtechnology, Ulm University, Ulm, 89081, Germany, where he earned his ``Habilitation'' (post-doctoral lecturing qualification) for Automation Technology in 2022. His research interests comprise connected automated driving, electric mobility, modelling and control of mechatronic systems, and system identification.
\end{IEEEbiography} \vspace{-15 mm}
\begin{IEEEbiography}[{\includegraphics[width=1in,height=1.25in,clip,keepaspectratio]{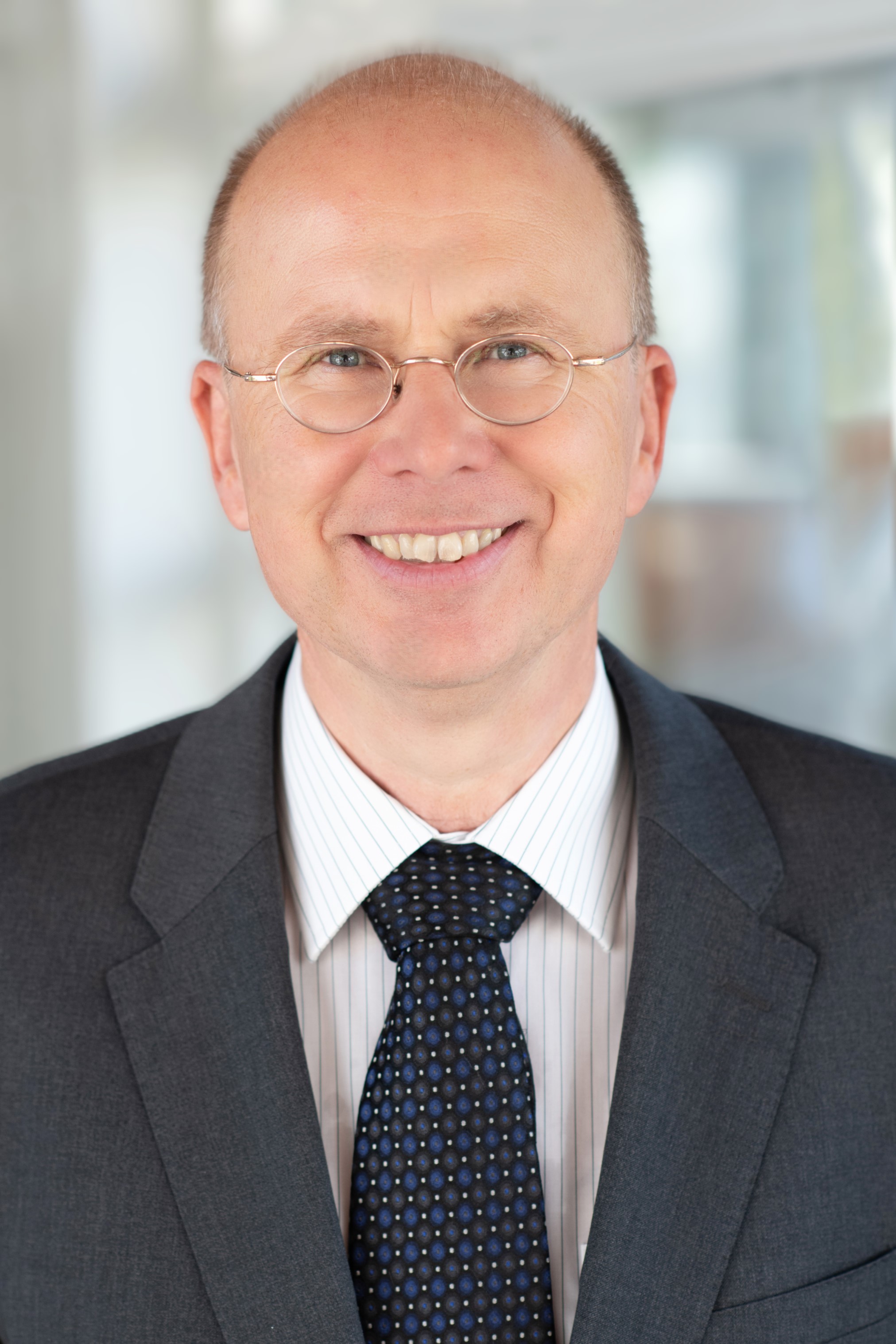}}]{Klaus Dietmayer}
	(Senior Member, IEEE) earned his degree in electrical engineering from the Technical University of Braunschweig, Germany and completed his Ph.D. in 1994 at the University of the Armed Forces, Hamburg, Germany. Afterwards, he began his industrial career as a research engineer at Philips Semiconductors, Hamburg, progressing through various roles to become the manager for sensors and actuators in the automotive electronics division. 
	
	In 2000, Dietmayer was appointed as a Professor of Measurement and Control at the University of Ulm. He currently serves as the Director of the Institute for Measurement, Control, and Microtechnology within the School of Engineering and Computer Science.
	
	His primary research interests include information fusion, multi-object tracking, environment perception, situation assessment, and behavior planning for autonomous driving. The institute operates three automated test vehicles with special licenses for public road traffic, along with a test intersection equipped with infrastructure sensors for evaluating automated and networked cooperative driving in Ulm.
\end{IEEEbiography} \vspace{-15 mm}
\begin{IEEEbiography}[{\includegraphics[width=1in,height=1.25in,clip,keepaspectratio]{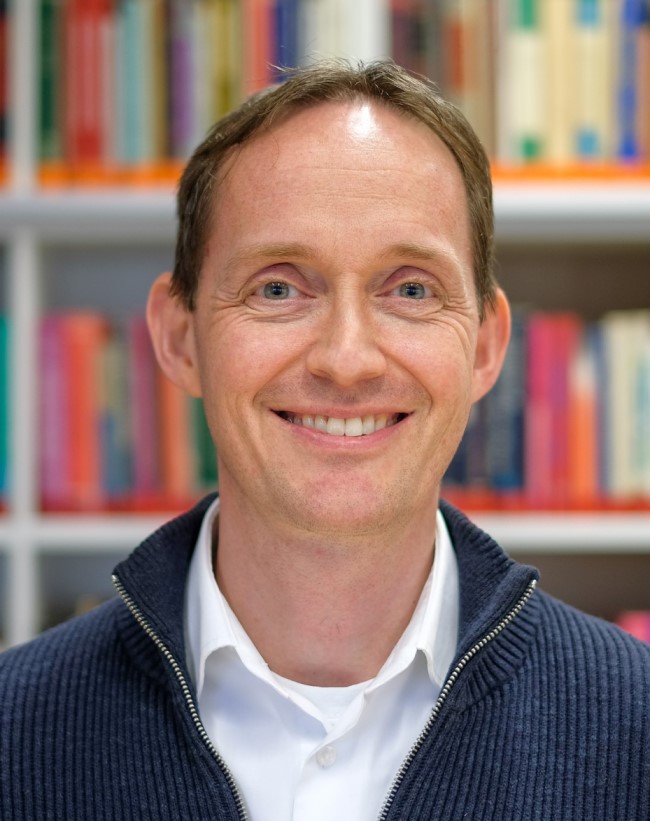}}]{Knut Graichen}
	(Senior Member, IEEE) received the Diploma-Ing. degree in engineering cybernetics and the Ph.D. (Dr.-Ing.) degree from the University of Stuttgart, Stuttgart, Germany, in 2002 and 2006, respectively.
	
	In 2007, he was a Post-Doctoral Researcher with the Center Automatique et Syst\`emes, MINES ParisTech, France. In 2008, he joined the Automation and Control Institute, Vienna University of Technology, Vienna, Austria, as a Senior Researcher. In 2010, he became a Professor with the Institute of Measurement, Control and Microtechnology, Ulm University, Ulm, Germany. Since 2019, he has been the Head of the Chair of Automatic Control, Friedrich–Alexander–Universita\"at Erlangen–N\"urnberg, Germany. His current research interests include distributed and learning control and model predictive control of dynamical systems for automotive, mechatronic, and robotic applications.
	
	Dr. Graichen is the Editor-in-Chief of Control Engineering Practice.
\end{IEEEbiography}\newpage

%

%

\end{document}

%% file: settings/packages.tex
\usepackage{amsmath,amsfonts,amssymb}
\usepackage{algorithmic}
\usepackage{algorithm}
\usepackage{array}
\usepackage[caption=false,font=normalsize,labelfont=sf,textfont=sf]{subfig}
\usepackage{textcomp}
\usepackage{stfloats}
\usepackage{url}
\usepackage{verbatim}
\usepackage{graphicx}
\usepackage{cite}

\usepackage{xcolor}
\usepackage{siunitx}
\usepackage{booktabs}

\usepackage{xurl}
\usepackage{tabularx}
\usepackage{diagbox}
\usepackage{makecell}

\usepackage{tikz}

\usepackage{bbding}
\usepackage{pifont}
\newcommand{\cmark}{\ding{51}}%
\newcommand{\xmark}{\ding{55}}%
\usepackage{multirow}

\usepackage{scalerel}
\usepackage{tikz}
\usetikzlibrary{svg.path}

\definecolor{orcidlogocol}{HTML}{A6CE39}
\tikzset{
	orcidlogo/.pic={
		\fill[orcidlogocol] svg{M256,128c0,70.7-57.3,128-128,128C57.3,256,0,198.7,0,128C0,57.3,57.3,0,128,0C198.7,0,256,57.3,256,128z};
		\fill[white] svg{M86.3,186.2H70.9V79.1h15.4v48.4V186.2z}
		svg{M108.9,79.1h41.6c39.6,0,57,28.3,57,53.6c0,27.5-21.5,53.6-56.8,53.6h-41.8V79.1z M124.3,172.4h24.5c34.9,0,42.9-26.5,42.9-39.7c0-21.5-13.7-39.7-43.7-39.7h-23.7V172.4z}
		svg{M88.7,56.8c0,5.5-4.5,10.1-10.1,10.1c-5.6,0-10.1-4.6-10.1-10.1c0-5.6,4.5-10.1,10.1-10.1C84.2,46.7,88.7,51.3,88.7,56.8z};
	}
}

\newcommand\orcidicon[1]{\href{https://orcid.org/#1}{\mbox{\scalerel*{
				\begin{tikzpicture}[yscale=-1,transform shape]
					\pic{orcidlogo};
				\end{tikzpicture}
			}{|}}}}

\usepackage{hyperref}

%% file: settings/commands.tex
\title{Expanding the Classical V-model to Cover Emerging Complex Systems Incorporating AI*}

\title{Expanding the classical V-model for the development of complex systems incorporating AI*}

\author{Lars Ullrich \orcidicon{0009-0001-8166-3118} \IEEEauthorrefmark{2}, Michael Buchholz \orcidicon{0000-0001-5973-0794} \IEEEauthorrefmark{3}, Klaus Dietmayer \orcidicon{0000-0002-1651-014X} \IEEEauthorrefmark{3},~\IEEEmembership{Senior Member,~IEEE,} \\and Knut Graichen \orcidicon{0000-0003-2865-8093} \IEEEauthorrefmark{2},~\IEEEmembership{Senior Member,~IEEE}
	\thanks{*This research is accomplished within the project ”AUTOtechagil” (FKZ 01IS22088Y, FKZ 01IS22088W). We acknowledge the financial support for the project by the Federal Ministry of Education and Research of Germany (BMBF).}
	\thanks{\IEEEauthorrefmark{2}Chair of Automatic Control, Friedrich-Alexander-Universität Erlangen-Nürnberg (FAU), Cauerstraße 7, 91058 Erlangen, Germany {\tt\footnotesize \{lars.ullrich, knut.graichen\}@fau.de}}
	\thanks{\IEEEauthorrefmark{3}Institute of Measurement, Control and Microtechnology, Ulm University, Albert-Einstein-Allee 41, 89081 Ulm, Germany {\tt\footnotesize \{michael.buchholz, klaus.dietmayer\}@uni-ulm.de}}
	\thanks{Manuscript received April 19, 2021; revised August 16, 2021.}}

%% file: src/00_Abstract.tex
\begin{abstract}\label{00_Abstract}
Research in the field of automated vehicles, or more generally cognitive cyber-physical systems that operate in the real world, is leading to increasingly complex systems. Among other things, artificial intelligence enables an ever-increasing degree of autonomy. In this context, the V-model, which has served for decades as a process reference model of the system development lifecycle is reaching its limits. To the contrary, innovative processes and frameworks have been developed that take into account the characteristics of emerging autonomous systems. To bridge the gap and merge the different methodologies, we present an extension of the V-model for iterative data-based development processes that harmonizes and formalizes the existing methods towards a generic framework. The iterative approach allows for seamless integration of continuous system refinement. While the data-based approach constitutes the consideration of data-based development processes and formalizes the use of synthetic and real world data. In this way, formalizing the process of development, verification, validation, and continuous integration contributes to ensuring the safety of emerging complex systems that incorporate AI. 
\end{abstract}


%% file: src/01_Introduction.tex
\section{Introduction}\label{01_Introduction}
\IEEEPARstart{T}{he} V-model \cite{brohl1993v} has served as a valuable tool for the safe and reliable development, verification, validation, and introduction of technical systems in the past. However, it is only suitable to a limited extent for the complex systems that are emerging today and will be in the future. In addition to the generally increased complexity of systems, the growing prevalence of artificial intelligence (AI) poses a particular challenge, even for more recent versions of the V-model \cite{graessler2018v}, \cite{GraesslerHentze}.

AI offers the advantage of mapping the system behavior without explicitly extracting and modeling relationships, instead emulating the input-output behavior based on data. This characteristic, along with technological advancements in artificial intelligence, holds promise for achieving higher levels of autonomy, which is why AI is increasingly being incorporated. At the same time, the transition from mathematically-explicit systems to data-based implicit systems present several challenges \cite{amodei2016concrete}, \cite{neto2022safety}, especially in safety-critical systems \cite{kurd2007developing, forsberg2020challenges} like automated driving where the provability of proper functionality is crucial and required \cite{europeancommissionaiact}. To address these challenges, the use of real world as well as systematically generated synthetic simulation data becomes indispensable in the development \cite{KIDeltaSynData} and approval processes \cite{KIAbsicherungSynData}, ensuring both safety and cost-effectiveness. This necessitates a process realignment to effectively consider and use various types of data.

Especially the need for both economic efficiency and safety urges the development of new innovative process reference models for the product development lifecycle alongside with continuous integration as well as addressing safeguarding and approval. This has initiated numerous research projects dealing with these crucial aspects, particularly in the field of automated driving \cite{PATH, SAKURA, V4SAFETY, PEGASUS, HIDrive, KIFamilie, LOPAAS, VIVID, SUNRISE}. In parallel, also the industry is actively engaged in addressing relevant topics to enable the approval of their systems \cite{karpathy_cvpr21, favaro2023building}. 

This paper is intended to link and bridge the different perspectives and methodologies by highlighting their commonalities and ultimately formalizing a unifying process reference model. Accordingly, the main contribution is threefold:

\begin{itemize}
	\item We present an analysis of existing development processes along with a comparison with regard to complex systems incorporating AI and a subsequent discussion of the current state of the art.
	\item We propose a formalizing, harmonizing, and generalizing framework entitled "iterative data-based V-model" expanding the classical V-model \cite{brohl1993v} for the development of complex systems that include AI.
	\item To illustrate the use of the proposed framework without exceeding the scope, we sketch the application of the concept to the use case of automated driving at different levels of granularity.
\end{itemize}

Accordingly, it should be emphasized that this paper is not a comprehensive description of a process that includes a fully-fledged safety assurance directly leading to approval. Instead, the purpose of this paper is to introduce a formalized process reference model for the entire product lifecycle at a macro level that aims to establish a general framework for the development, verification, validation, and continuous integration of complex systems incorporating AI. Specifically, the proposed iterative, data-based V-model is intended to represent a generic and structured process, similar to the classical V-model \cite{brohl1993v}, that describes the process but leaves the safety argumentation and assessment custimizable in order to achieve the desired generality and transferability, ultimately providing an expanded guidance for complex systems featuring AI.

The paper is structured as follows: The related work outlines relevant research projects in Section \ref{02_StateOfTheArt}. Building on this, the analysis of innovative development processes in Section \ref{02_New} investigates a selection of methods in relation to the classical V-model and compares them concerning complex systems. The methodology of the iterative data-based V-model is elaborated, classified, delimited and summarized in Section \ref{03_Methodology}. The usability and application of the proposed methodology are abstractly sketched in Section \ref{04_Examples} using an academic use case of automated driving. Finally, the reference process model presented is discussed and conclusions are drawn in Section \ref{05_Conclusion}.

%% file: src/02_StateOfTheArt.tex
\section{Related Work}\label{02_StateOfTheArt}
This section provides a general overview of relevant research projects and highlights a selection that is to be examined in more detail in the subsequent section.  

The Pegasus family \cite{PEGASUS} comprises the Pegasus \cite{PEGASUS} project as well as the successor projects Verification Validation Methods (VVM) \cite{VVM} and SetLevel \cite{SETLevel}. While the SetLevel project focused on the particular challenge of simulation-based development, the VVM project built upon the decomposition of the operational design domain (ODD) into logical core scenarios established by the Pegasus project and extends this methodology in a holistic manner. Consequently, the overall methodology of VVM \cite{VVMOverall} represents an extension of the traditional V-model. A similar approach is being pursued by the Japan Automobile Manufacturers Association (JAMA) with its Automated Driving Safety Evaluation Framework \cite{JAMAFramework}. This framework also constitutes a scenario-based approach and an enhanced V-model structure. The backbone of these scenario-based approaches are scenario databases, for instance from the ADScene \cite{guyonvarch2023adscene}, Safety Pool \cite{SafetyPool}, and SAKURA \cite{SAKURA} projects. Moreover, the similarity in the approaches results in part from close exchange and cooperation. For example, as part of the German-Japanese VIVID \cite{VIVID} project, the national sub-projects VIVALDI and DIVP \cite{DIVP} were conducting joint research on virtual validation. 

In the context of scenario-based approaches, the theoretical framework of Scenarios Engineering (SE) \cite{li2022novel, li2022features} is also worth mentioning. This framework aims to improve the visibility, interpretability, and reliability of intelligent systems and, thus, contributes to the realization of trustworthy AI. Beyond this, the emergence of video generation models, such as \mbox{OpenAI's} Sora, is addressed in \cite{li2024sora} as a step towards imaginative intelligence, while discussing its relevance in the context of SE. As outlined, the advancement supports the training and testing of intelligent vehicles by reducing the need for physical recordings \cite{li2023novel} and expanding the variety of scenarios. In addition, \cite{wang2024does} elaborates on the opportunities that imaginative intelligence offers for SE, including tackling the long-tail problem, while also outlining the challenges that currently remain in accurately modeling physics and understanding causality. Thus, \cite{li2024sora} and \cite{wang2024does} take into account recent developments and present a perspective of future developments. In this regard, the NXT GEN AI METHODS \cite{NextAIM} project should also be noted, which is dedicated to generative AI and, more specifically, the development of foundation models for automated driving.

Furthermore, there are several projects like StreetWise \cite{elrofai2018streetwise}, HEADSTART \cite{HEADSTART}, SAFE-UP \cite{SAFEUP}, and AI Safeguarding \cite{ki_absicherung} that focused on validation, safety assessment, and safety assurance within the field of automated driving. Moreover, projects such as V4SAFETY \cite{V4SAFETY} and SUNRISE \cite{SUNRISE} aim to develop more comprehensive frameworks for safety assessment and assurance, respectively. The LOPAAS \cite{LOPAAS} project between Fraunhofer IESE, Fraunhofer IKS, and the Assuring Autonomy International Program (AAIP) at the University of York also targets to bring about a paradigm shift in safety engineering for autonomous systems. Hi-Drive \cite{HIDrive}, another project, addresses the various ODD challenges but concentrates on reaching a higher level of autonomy. Consequently, a multitude of projects is anchored in this context. 

However, the VVM \cite{VVM} project is of particular interest as it can be considered the most advanced representative of the improved V-model approaches and outlines an overall methodology. Of particular interest is also the methodology of Waymo \cite{favaro2023building}, which differs in focus from the VVM \cite{VVM} project. In addition to the different focus, the methodology is of particular interest due to the resulting comparatively solid performance \cite{kusano2023comparison} of the automated vehicles in comparison to other competitors \cite{equipmentrecallreport, NHTSARecall23E, NHTSARecallLetter}. Beyond that, Tesla's approach \cite{karpathy_cvpr21} is technologically different and benefits from a large number of vehicles in the field that generate data. As a consequence, the methodology is based on an AI-centric development process, similar to processes of other AI systems such as OpenAI's ChatGPT. For this reason, the general approach, also known as the data engine is also worth emphasizing. Ultimately, the three highlighted processes are to be analyzed in more detail subsequently.

\section{Analysis of Innovative Development Processes}\label{02_New}
While the previous section provides a general overview, this section is dedicated to the analysis of a selection of innovative development processes. Thereby, the different advancements and perspectives are related to the classical V-model in order to set a basis for the subsequent creation of a generalized framework. For this purpose, the VVM \cite{VVM} project, Waymo's methodology \cite{favaro2023building} and Tesla's approach \cite{karpathy_cvpr21} are examined in more detail below. In addition, the methodologies are compared in terms of their general applicability to complex systems. Conclusively, the application areas of the available frameworks and the existing gap in the overall context are discussed.  

\subsection{VVM Project: Argumentation- \& Scenario-based V-model}\label{VVM}

The VVM project enhances the traditional V-model \cite{brohl1993v} by integrating multiple perspectives. First, it employs the ODD decomposition from Pegasus \cite{schuldt2013effiziente, pegasus_schlussbericht} to construct an ODD metamodel \cite{scholtes20216, reich2023concept}, facilitating scenario-based design and Verification \& Validation (V\&V) processes \cite{elster2021fundamental}. This method effectively maps infinite scenarios of the open world into manageable test spaces and design foundations \cite{neurohr2021criticality}. Second, a coherent assurance argumentation \cite{VVMAssurance} targets the mitigation of unreasonable risks in the open world, fostering consistency and traceability along the V-model. Consequently, the project establishes consistent interfaces across the framework and enables seamless requirements considerations from design to verification via argumentation-based V\&V. Third, a multi-perspective approach \cite{VVMAPerspectives} addresses systematic gaps such as specification, implementation, and validation gaps, aiming at uncertainty and risk reduction \cite{stellet2019formalisation}. This involves decomposing the overall system into various levels of abstraction, incorporating capability, engineering, and real-world layers, and employing different perspectives like design, V\&V, risk management, and argumentation.

\begin{figure}[]
	\centering	
	\includegraphics[width=\linewidth]{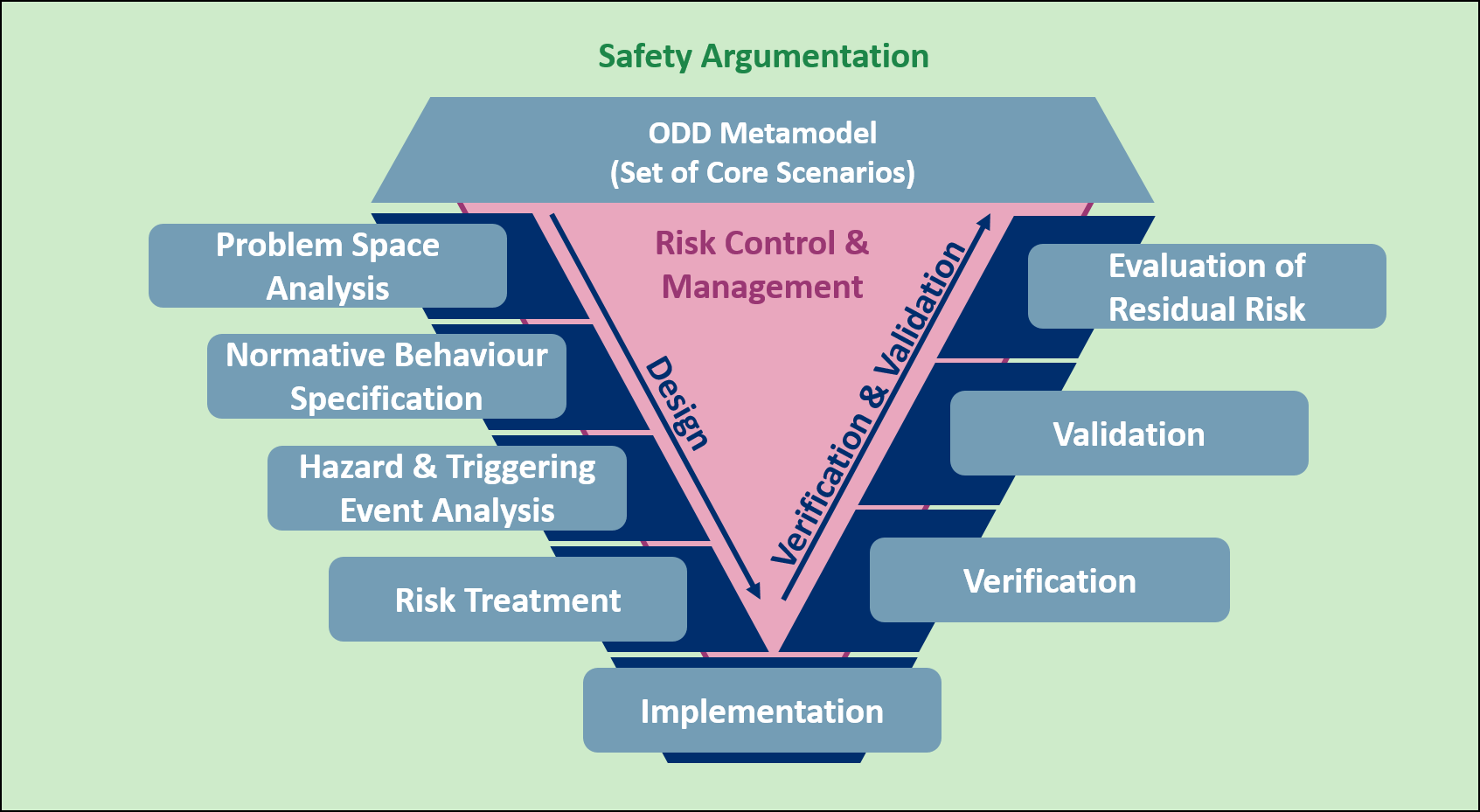}
	\caption{Visualization of the overall methodology of the VVM project \cite{VVMOverall} as an extension of the classical V-model designed for automated driving application with regard to a scenario-based problem decomposition and an appropriate safety argumentation.}
	\label{fig:VVM}
\end{figure}

Finally, the overall methodology culminates to an advanced V-model \cite{VVMOverall}, shown in Figure \ref{fig:VVM}, that is customized to the scenario-based logic introduced by Pegasus. 

The overall methodology culminates in an advanced V-model \cite{VVMOverall}, as depicted in Figure \ref{fig:VVM}, customized to the scenario-based logic introduced by Pegasus. The ODD metamodel forms the top of the model, inducing the scenario-based philosophy throughout the V-model process. Problem space analysis provides the basis for detailed specification, considering the environment and the specific automated driving system. Furthermore, to ensure regulatory compliance, normative behavior \cite{salem2022beitrag} is specified, encompassing certification, legal, social, and ethical expectations. Subsequently, systematic hazard  \cite{graubohm2020towards} and risk identification are conducted based on the ODD metamodel \cite{VVMOverall}, problem space analysis, and normative behavior specification. Corresponding safety measures to determine an acceptable residual risk are then defined through risk treatment, closely coordinated with the underlying sub-framework, the risk management core \cite{salem2023risk}.

After implementation, a scenario-based verification, validation, and risk assessment is performed \cite{riedmaier2020survey}. This process incorporates three key improvements over the classical V-model: access to the ODD metamodel for aligned analysis, assurance assessment throughout V\&V, and evaluation of residual risk \cite{VVMOverall}. In summary, the VVM project provides an extended, detailed, and tailored process reference model for scenario-based development and V\&V of automated vehicles.

\subsection{Waymo: Safety Determination Lifecycle}\label{Waymo}

While Waymo's safety determination lifecycle also strives for the absence of unreasonable risk, it places a stronger emphasis on the lifecycle \cite{favaro2023building}. Furthermore, similar to the VVM project \cite{VVMAPerspectives, stellet2019formalisation}, the framework encompasses different perspectives. Overall, the framework can be summarized as a layered, credible, and dynamic approach \cite{favaro2023building}.

The \textit{\textbf{layered approach}} within \cite{favaro2023building} refers to a division into an architectural (formerly known as hardware \cite{webb2020waymo}), a behavioral, and an in-service operational layer. Thereby, in order to demonstrate the absence of unreasonable risk, hazards and appropriate acceptance criteria are defined within the aforementioned layers. This involves the definition of several dimensions of interest covering a diverse set of aspects, e.g. the avoidance of incidents, the successful completion of automated journeys or the compliance with driving rules. Indicators of interest are defined on this basis, which map hazards to an explicit set of acceptance criteria. Through the definition of the minimum dimensions of interest, Waymo determines the completeness of the set of acceptance criteria in order to underpin credibility.  

The \textit{\textbf{credible approach}} adresses concerns about the reasonableness and trustworthiness of the claim-argument-evidence structure via Waymo's novel Case Credibility Assessment (CCA) \cite{favaro2023building}. The CCA comprises three components: the top-down credibility of the argument, the bottom-up credibility of the evidence, and the encompassing implementation of the credibility. Overall, the CCA procedure comprises a continuous revision by monitoring and updating the arguments and evidence to achieve credibility.

The top-down credibility of the arguments focuses on fulfilling overarching objectives through assessing the suitability and reasonableness of the arguments. This involves evaluating and refining a collection of arguments and acceptance criteria. Additionally, it entails justifying acceptance criteria and conducting a suitability assessment of performance indicators and associated objectives to evaluate the reasonableness. In contrast, the bottom-up credibility of the evidence concentrates on evaluating the evidence provided by the methodology. This entails analyzing the evidence with regard to both technical engineering and process management to assess confidence. Additionally, it entails evaluating the representativeness and applicability of the evidence for coverage assessment.

The \textit{\textbf{dynamic approach}} within \cite{favaro2023building} emphasis the siginficance of continuous assessment and refinement. Thereby, the time frame across the itervative procedures is distinguished in three phases: the pre-deployment, the deployment, and the post-deployment phase \cite{favaro2023interpreting}. During pre-deployment, design and V\&V are prospective, with performance measures based on simulations and field operations. Accordingly, acceptance criteria are treated as predicted values until deployment. Successive post-deployment involves retrospective performance analysis in the open world \cite{scanlon2023benchmarks}. Continuous monitoring identifies gaps, challenges, threats, and hazards, addressing the dynamic challenges of the open world and facilitating ongoing refinement of the system and the process throughout the product lifecycle.

\begin{figure}[]
	\centering	
	\includegraphics[width=\linewidth]{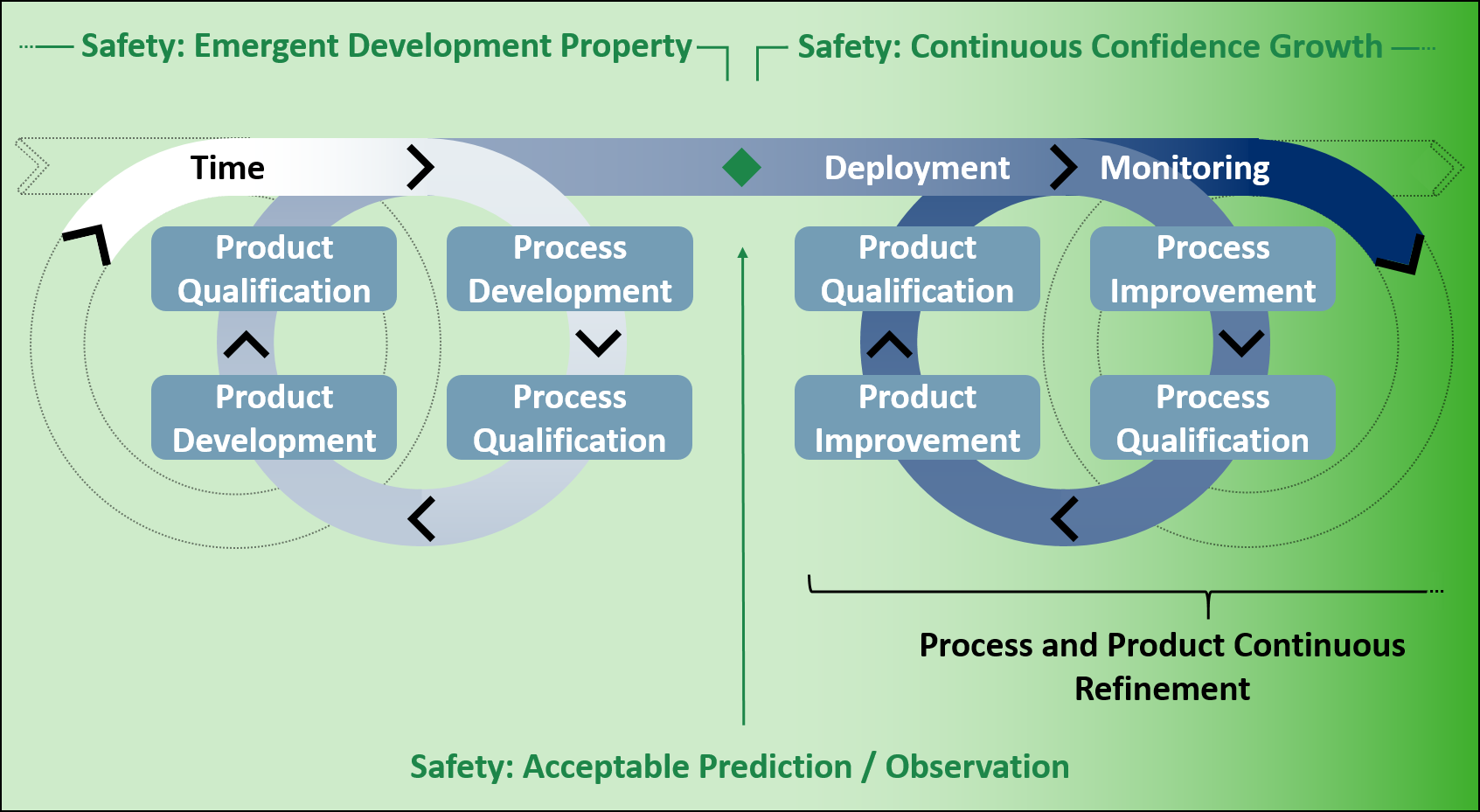}
	\caption{Simplified representation of Waymo's safety determination lifecycle, inspired by \cite{favaro2023building}. It illustrates the distinguished consideration of prospective and retrospective perspectives on the methodology and safety argumentation.}
	\label{fig:Waymo}
\end{figure}

The \textbf{overall methodology} of the safety determination lifecycle is depicted in Figure \ref{fig:Waymo}, emphasizing the dynamic approach. Thereby, in accordance with the three phases of the dynamic approach, safety evolves as an emergent property, an acceptable prediction, and a constantly growing confidence. Morover, to achieve the desired safety and credibility, Waymo considers the process and the product to be aligned. Each iteration consists of process and product development/refinement and subsequent qualification, reflecting two consecutive V-models, one for the process and another for the product. The approach addresses the complexities of system development and real-world applicability through ongoing refinement, effectively integrating development and analysis. Over time, as the system scales, uncertainty diminishes, confidence increases, and real-world performance is evaluated against human benchmarks \cite{di2023comparative, kusano2023comparison}.

\subsection{Tesla: Data Engine}\label{Tesla}

\begin{figure}[]
	\centering	
	\includegraphics[width=\linewidth]{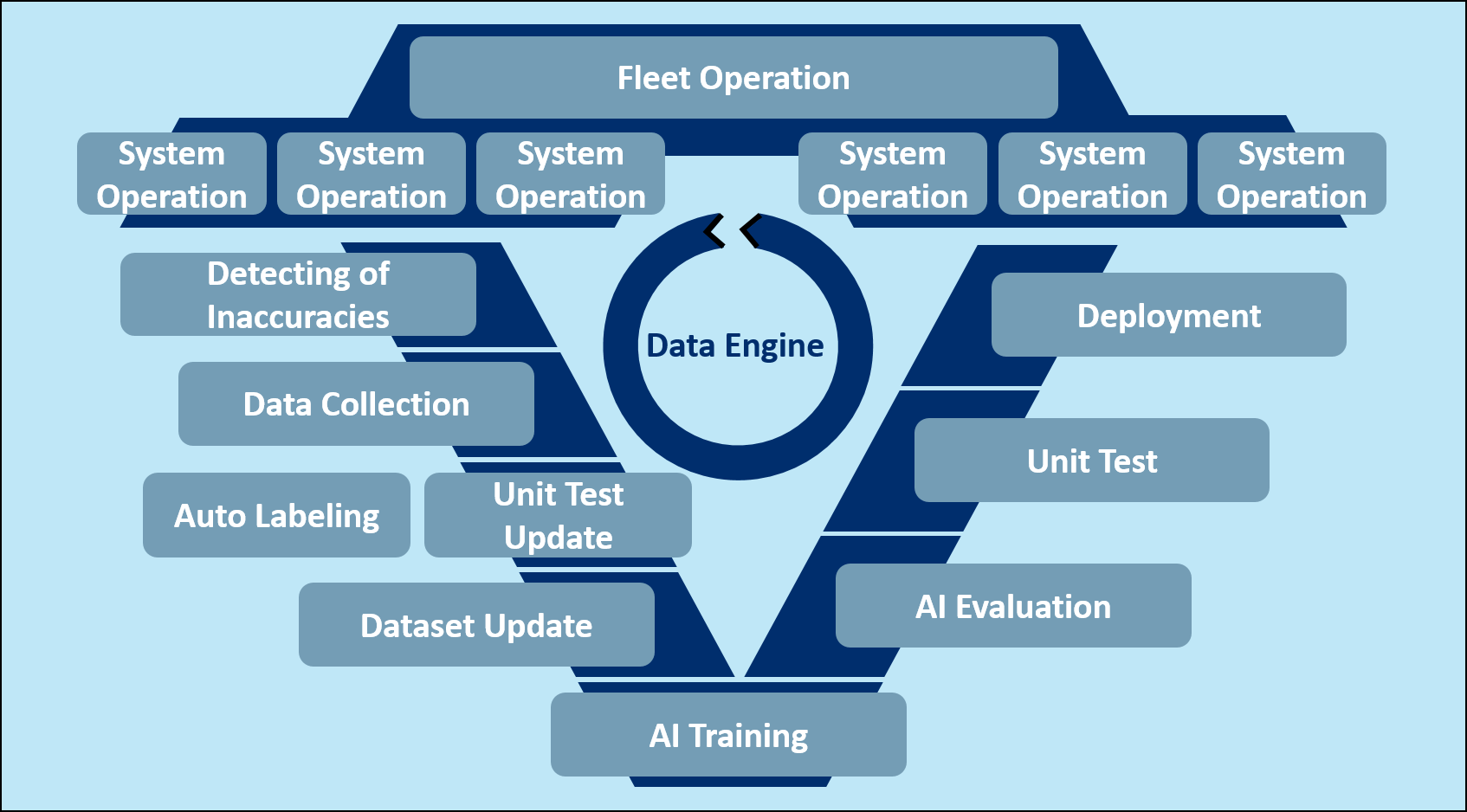}
	\caption{Tesla's data engine \cite{karpathy_cvpr21}, visualized in V-model structure, represents a fully data-driven methodology that is tailored to AI systems and strives for efficient and effective continuous improvement.}
	\label{fig:Tesla}
\end{figure}

The functional objectives of Tesla's Full Self-Driving and Waymo's self-driving service, Waymo One, are comparable, but they employ different technological and methodological approaches. Tesla relies on an end-to-end AI strategy based on camera data and benefits from a large fleet of operational vehicles, providing a vast amount of data for development and V\&V. Tesla's corresponding methodological approach, known as the data engine \cite{karpathy_cvpr21}, is illustrated in Figure \ref{fig:Tesla}. 

The data engine employs a data-centric, iterative development approach applicable to variety of AI systems and applications. Initially, an AI model is trained with a seed dataset. Initially, an AI is trained with a seed dataset. Subsequently, as in the case of Tesla, it is deployed in shadow mode in the customer's vehicles \cite{Tesla_shadow}, also known as silent testing \cite{templeton2019}. This involves employing specialized mechanisms to detect neural network inaccuracies, facilitating strategic data acquisition. Tesla, for instance, has crafted over 200 triggers to detect discrepancies while predicting surrounding object parameters like position, velocity, and acceleration \cite{karpathy_cvpr21}. While these triggers can be seen as AI performance indicators with respect to Waymo's terminology \cite{favaro2023building}, they serve a different purpose in this context.

The data engine's collection process drives follow-up auto-labeling and unit test updates, ensuring continuous dataset refinement. Subsequently, the neural network undergoes retraining, evaluation, and unit tests to avoid inaccuracies in subsequent versions, before redeployment in shadow mode \cite{Tesla_shadow, karpathy_cvpr21}. This systematic approach resembles a sophisticated automated data-based system, resembling a scenario-based database, effectively addressing real-world gaps and changes over time. However, for a seamless transition from silent testing to actual operation \cite{Tesla_shadow}, corresponding safety argumentation and assurance are essential. Nevertheless, Tesla does not provide any further information on this \cite{tesla_safety}.

The data engine not only enables systematic data acquisition and continuous improvement but also provides additional benefits for developing AI-based systems with large datasets. This includes increased efficiency through automation, creating a consistent database, and avoiding redundant data, leading to cost savings and improved effectiveness for diverse engineering teams.

Moreover, investigations reveal the widespread adoption of the data engine process, exemplified by its use by OpenAI \cite{DALLE2}. This indicates its versatility in addressing specifications for inadmissible outputs by selectively filtering undesirable data before training. Serving as an active learning approach, the data engine process collects data, updates AI systems, and mitigates the risk of catastrophic forgetting through offline updates, including V\&V and unit testing, before deployment. Therefore, the selection as a data-based process reference model seems appropriate.

\subsection{Comparison w.r.t. Complex Systems Incorporating AI}\label{sec:diff}

While previously the selected innovative development processes have been presented, these are now compared below in terms of general applicability to complex systems that incorporate AI. Emerging complex systems, like automated driving, feature increasing AI integration alongside traditional systems and heightened task complexity, leading to intricate architectures and necessitating complex orchestrations. The comparison of these processes in terms of their key advantages and disadvantages for general complex systems is presented in Table \ref{tab:compare_frameworks_ad_disad}. However, inherent limitations arise concerning their applicability to complex systems with AI integration due to application specific assumptions and conditions.

\begin{table*}[]
	\centering
	\caption{Comparison of development processes for complex systems incorporating AI.}
	\begin{tabularx}{\linewidth}{l *{2}{>{\raggedright\arraybackslash}X}}
		\toprule
		Framework	& \makecell{Advantages} & \makecell{Disadvantages}  \\
		\midrule
		\makecell[l]{Improved V-model \\ \tiny{(VVM Project)}} \vspace*{-0.4cm} &  \vspace*{-0.4cm} \begin{itemize}
			\item Systematic improvement of the classical V-model
			\item Consitent consideration of the operational design domain
			\item Maps infinite scenarios of the open world
			into a managable test space
		\end{itemize} \vspace*{-0.4cm} &  \vspace*{-0.4cm}
		\begin{itemize}
			\item Application-specific for automated driving 
			\item Requires a scenario database, not suitable for general databases
			\item Requires multiple perspectives, e.g. design, V\&V, risk management, argumentation
			\item Lacks alignment with the requirements of AI systems
		\end{itemize} \vspace*{-0.4cm}  \\
		\midrule
		\makecell[l]{Safety Determination Lifecycle \\ \tiny{(Waymo)}} \vspace*{-0.4cm} &  \vspace*{-0.4cm}
		\begin{itemize}
			\item Consideration of prospective and retrospective perspectives
			\item Alignment and refinement of process and product throughout the entire lifecycle
			\item Strong emphasis on V\&V and approval process, e.g. through case credibility assessment
		\end{itemize} \vspace*{-0.4cm} &  \vspace*{-0.4cm}
		\begin{itemize}
			\item Tailord towards the automated driving application
			\item Requires several approaches, e.g. layered, credible, dynamic
			\item Less formalized development process, due to strong emphasis on V\&V and approval process
		\end{itemize}  \\
		\midrule
		\makecell[l]{Data Engine \\ \tiny{(Tesla)}} \vspace*{-0.4cm} &  \vspace*{-0.4cm}
		\begin{itemize}
			\item Considers specifics of data-driven AI development and V\&V 
			\item Able to be automated, reduces human efforts and influences
			\item Provides application-agnosticity and iterative refinement
		\end{itemize} \vspace*{-0.4cm} &  \vspace*{-0.4cm}
		\begin{itemize}
			\item Not applicable to traditional or mixed systems
			\item Requires systems in operation for silent testing and data acquisition
			\item Less sophisticated process, e.g no simulation, no process refinement
		\end{itemize}  \vspace*{-0.4cm} \\
		\bottomrule
	\end{tabularx}%
	\label{tab:compare_frameworks_ad_disad}
\end{table*}%

The VVM project's improved V-model \cite{VVMOverall} and Waymo's safety determination lifecycle \cite{favaro2023building} embrace multiple perspectives, incorporate system structures, address different granularities, and onsider the safety argumentation, whereas the classic V-model \cite{brohl1993v} takes a singular approach focusing on design, V\&V, and customizable safety argumentation. This distinction allows the classical V-model to maintain broad applicability across various systems by accommodating flexibility in architecture, granularity, and assurance methods to meet specific requirements. These properties are essential for a framework aiming to address the needs of diverse complex systems. 

Moreover, Tesla's data engine \cite{karpathy_cvpr21} is primarily designed for AI systems, making it less adaptable to traditional or mixed systems. Moreover, it relies on a large-scale human-operated system fleet for data collection and silent testing, a requirement that is often impractical

In comparison to partially extended responsibilities and non-generic assumptions, frameworks often overlook formalizing the interdependence and transitions between simulation and the real world. However, it is evident that the joint interaction between simulation and the real world  is pivotal for complex systems \cite{KIDeltaSynData, KIAbsicherungSynData}. This interaction can result in significant gaps \cite{burton2023closing, burton2023addressing} during both the design and V\&V phases. Currently, there's a lack of a systematic formalization for the transition from the real world to simulation and back within a continuous refinement process. Such a formalization that address concerns related to data-driven development without excluding traditional and mixed systems, while ensuring flexibility in terms of safety assessments and argumentation, does not yet exist. 

\subsection{Discussion}
\begin{figure}[h!]
	\centering	
	\includegraphics[scale=0.45]{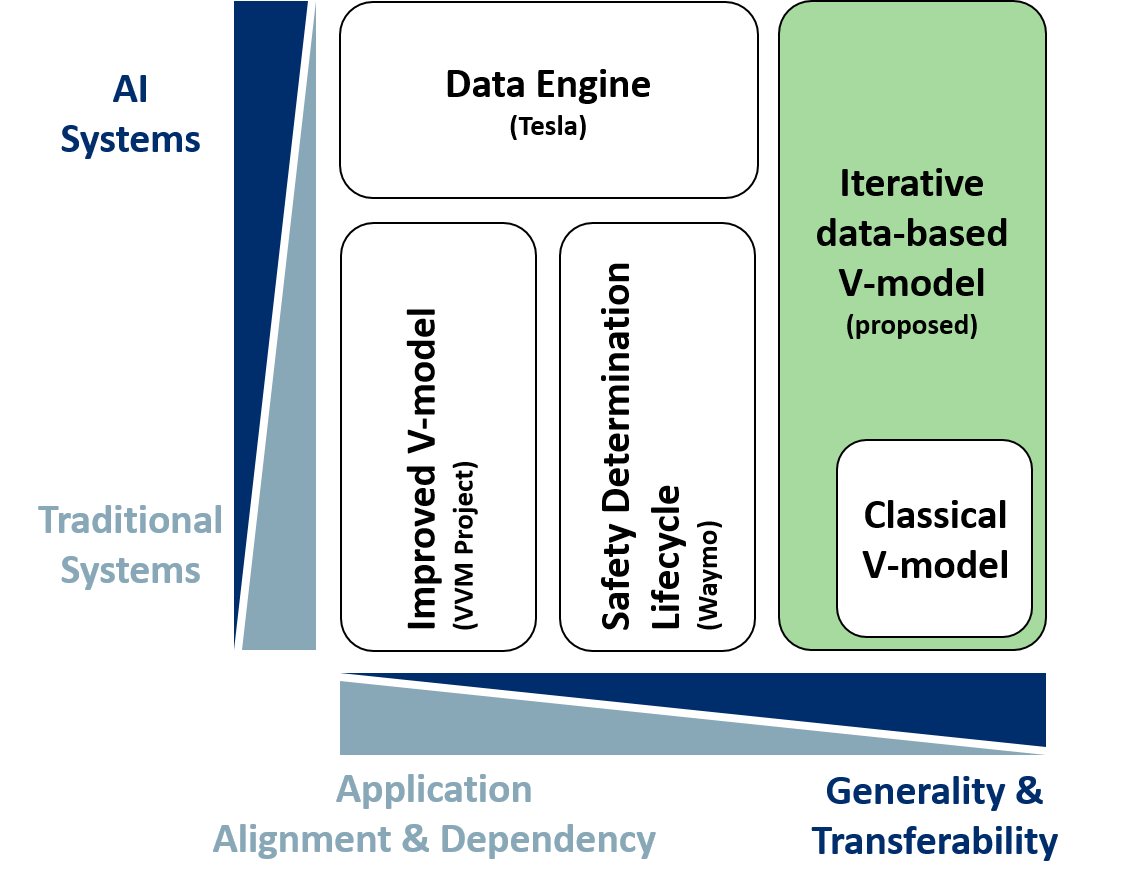}
	\caption{Visualization of the classical V-model and the analyzed innovative development processes with regard to methodological generality and system compatibility. The remaining gap is marked by the proposed iterative data-based V-model.}
	\label{fig:cmp}
\end{figure}

The analysis of selected processes reveals that alongside the classical V-model, designed for traditional systems, various process reference frameworks have emerged to address specific complex systems integrating both traditional and AI components, which can be described as emerging complex systems that incorporate AI. The improved V-model of the VVM project represents an application-specific extension of the classical V-model and addresses complex systems of automated driving. Waymo's safety determination lifecycle is similarly motivated but less dependent on the application and system design. In comparison, Tesla's data engine explicitly addresses the characteristics of AI systems. Finally, Figure \ref{fig:cmp} visually compares the above process reference model frameworks in the tension between application specificity and generalization, and suitability for traditional systems, AI systems, and emerging complex systems that incorporate AI. On an abstract level, Figure \ref{fig:cmp} shows that each approach has a certain area of application. Furthermore, it becomes apparent that there is a gap in the area of generic processes for complex systems that incorporate AI. This ultimately motivates our contribution in visual form. The correspondingly proposed iterative data-based V-model, which is intended to narrow this gap, is elaborated in the following section.  
 
%% file: src/03_Methodology.tex
\section{Iterative Data-based V-model}\label{03_Methodology}

Even though the existing development and V\&V frameworks for emerging complex systems, particularly automated vehicles, differ in their terminology, methodological specifics, and different emphasis, there is a general common sense that arises from the inherent nature of the engineering processes. Furthermore, while corresponding safety assurance is crucial for approval, it is highly dependent on the system and the environment, which is also referred to as ODD. Therefore, the overall methodology should be decoupled from the application-specific safety assurance, generalized across existing frameworks, and formalized along the transitions between simulation and the real world, while striving to support and enable safety argumentation along the resulting iterative process reference model in a general form. This requires, similar to the classical V-model, a certain level of abstraction for widespread applicability. Therefore, a detailed, fully-fledged safety case that leads to direct release cannot be provided. Consequently, the iterative data-based V-model is introduced to build on the generality of the classical V-model, bridging existing frameworks while simultaneously addressing the challenges of emerging systems and technologies in a sophisticated manner. Thereby, a particular emphasis is laid on cognitive cyber-physical systems of the real world that are safety critical to adress challenges of arising autonomous technologies.

\subsection{Fundamental Principles}
Adressing emerging autonomous technologies in a general manner highlights the criticality of a generic scneario-based approach \cite{riedmaier2020survey, elster2021fundamental, VVMOverall}, as corresponding database generation could be exhaustive and costly. Consequently, this would limit the suitability of the methodology towards solid technological applications that justify the efforts. Elsewise, a silent testing \cite{Tesla_shadow, templeton2019} approach for corresponding data acquisition, as scenario database equivalent, is also limited by the fact that dedicated systems and applications must already be operating on a large scale in the target environment. Beyond that, direct data collection in the real world might also be inappropriate, as it could lead to hazards and does not provide the necessary scale of datasets. As a consequence, the general methodology of the emerging autonomous systems is based on systematically generated data on a large scale through simulation. This observation is inline with the project series AI Family \cite{KIFamilie} that covers a range of sub-projects on AI assurance \cite{KIAbsicherungSynData}, transfer \& scaling \cite{KIDeltaSynData}, data tooling \cite{KIDataTooling}, and the hybridization of knowledge and data for automated driving applications \cite{KIWissen_D1_D4, KIWissen_D2}. Although the project series dealt with core topics such as the exploitation of simulation and real data, an overarching reference process has not been established. 

\subsection{General Methodology}\label{sec:methodo}
\begin{figure*}[]
	\centering	
	\includegraphics[width=0.7\linewidth]{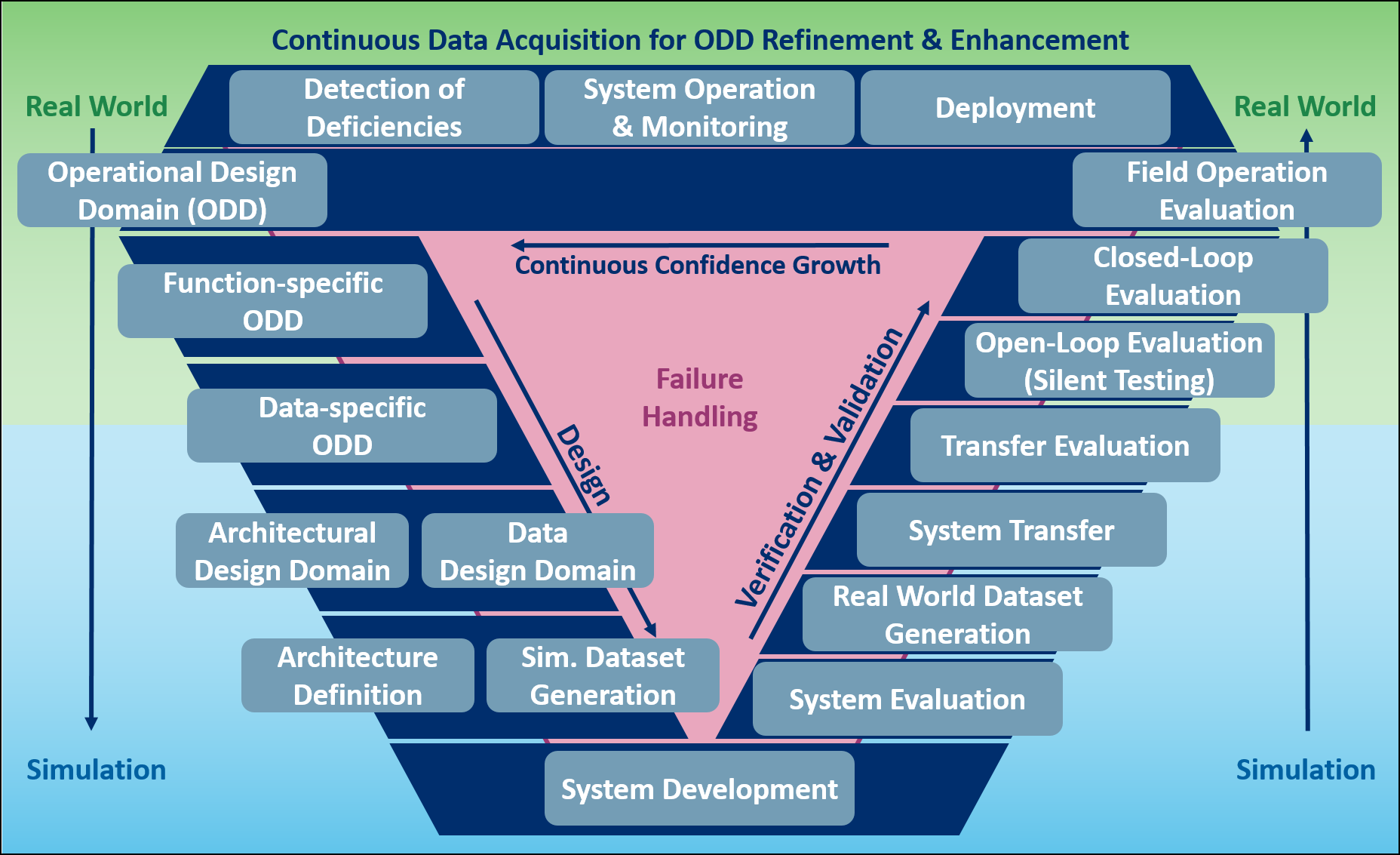}
	\caption{The iterative data-based V-model, which formalizes and merges the various existing methods. The initial loop starts with the definition of the ODD. The explicit formalization of the process from the real world to simulation and back and the data-based characteristic address the challenges of complex systems that embrace AI. The iterative approach, on the other hand, addresses the challenges of open world complexity and offers continuous system and confidence improvement in an intuitive way.}
	\label{fig:ours}
\end{figure*}

The proposed iterative data-based V-model builds on the VVM \cite{VVM} project by taking up the general V-model structure that is enhanced by a consistent alignment of the ODD along the V-stages. Beyond the application of automated driving, the ODD is interpreted as a generic concept that systematically defines the environment and context of the respective system. In addition, the iterative data-based V-model reflects Waymo's dynamic lifecycle approach \cite{favaro2023building}. In this way, the open world associated challenges as well as systematic gaps are addressed in a natural, iterative and continuous refinement manner. While the approach focuses on the product and provides guidance through the reference process, the chosen level of abstraction also allows for underlying process refinement along the safety argumentation. In addition, the iterative data-based V-model generalizes the scenario-based database approach. Furthermore, the iterative data-based V-model considers AI-specific development aspects and takes up ideas of the data engine \cite{karpathy_cvpr21} to increase efficiency. However, a central part remains a V\&V that is adapted to the challenge of the targeted use of simulation and the real world. Along a systematic consideration of the strengths of innovative frameworks while addressing existing limitations when applied to complex systems, as discussed in Section (\ref{sec:diff}). This results in an iterative data-based V-model, which is outlined in Figure \ref{fig:ours}. The individual aspects are described in more detail below. 

\textbf{Operational Design Domain:} The cycle initially starts with the definition of the ODD. Here, the ODD states an application independent foundation for defining the system's context, e.g. the environment and operating conditions. Besides that, the ODD represents the top-level target designation and belonging requirements specification. Moreover, the ODD provides guidance for the requirement definition of subordinate systems and functions. Thus, taking into account the system ODD independently of the granularity of the functionality or component to be developed ensures top-level aligment.

\textbf{Function-specific ODD:} Here, the specific functionality of the system, subsystem, or component to be developed is specified by means of a dedicated ODD. Thereby, following the top-down decomposition, the relevant top-level targets of the ODD are stringently broken down and transferred to the respective function-specific ODD. Moreover, specifics such as the desired behavior of the functionality that does not result directly from the top-level targets is included. Accordingly, the explicit function-specific ODD compactly defines the requirements specification while simultaneously minimizing specification uncertainties \cite{burton2023closing} through its alignment with the ODD.

\textbf{Data-specific ODD:} In this context, the previously outlined requirements specification is transfered into a data perspective. In this way, the data-specific ODD accounts for the fact that the availability, quality, and nature of the data can have a direct impact on how the system performs the task at hand. Thus, the definition of the data-specific ODD is responsible
for transforming the function-specific ODD requirements into the data-based representation. Making this transformation explicit within the framework should minimize the systematic gaps. In particular, the fact that development and V\&V are data-driven highlights the importance of a precise translation of the formally defined specifications and requirements into a data-specific representation.

Consequently, the data-specific ODD, which is responsible for the syntactic transition into the data domain while preserving the semantics to meet required demands, constitutes an interface. In more detail, the data-specific ODD represents an handover area between system domain experts and funtion experts (e.g. AI experts), who use different terminologies (linguistic gap) and have different understanding (knowledge gap) due to different backgrounds (specialization gap). This entails the risk that the semantics of the requirements are not fully translated and usually leads to a specification gap. The framework aims to tackle and minimize this risk, which has already been described as specification uncertainties \cite{burton2023closing} that lead to specification insufficiencies \cite{burton2023addressing} and result in a semantic gap. The goal of the data-specific ODD is to enhance the efficiency and effectiveness of developing and ultimately safeguarding emerging complex systems that incorporate AI. This target-oriented addressing of the semantic gap can lead to overarching improved performance, reliability, and robustness.

\textbf{Architectural Design Domain / Data Design Domain:}  The two components of this stage address the transition to development. The architectural design domain defines a framework of possible architectures of the system to be developed on basis of the data-specific ODD. Thus, on the one hand, this step defines requirements for the architecture while on the other hand first assumptions of the design process that constitute from the data-specific ODD are made explicit. Thus, this stage represents a design abstrachtion layer for general descisions during the development and improvement processes.

In parallel, the definition of the data design domain transfers the requirements of the data-specific ODD into the requirements for
generating the data. However, the generation of data is not part of this stage. This deliberate separation of the definition of a design domain from the realization is an integral part of the framework and aims at the explicit separation, formulation, and documentation of requirements and assumptions while at the same time disclosing process related
uncertainties. In addition, this enables a decoupled validation of the design domains and the implementation via the continuous refinement process and indicates the need for action. For example, if deficiencies are identified during monitoring, it is possible to check whether the data design domain was valid and the problem arises from the incompleteness of the dataset in relation to it, or whether a refinement of the previous specifications and assumptions is necessary. 

\textbf{Architecture Definition / Sim. Dataset Generation:} This stage represents a deliberated engineering stage that focus on both the architecture setup as well as the simulation based data generation. The architectural definition selects a certain architectur within the corresponding architectural design domain and determines the specifics like parameters. 

The simulation data generation characterizes the systematic synthesis of data with respect to the data design domain. While the architecture selects a dedicated realization from the set of possible architectures, the creation of the dataset aims at completeness. In line with other frameworks like \cite{VVMOverall, favaro2023building, karpathy_cvpr21} and standards such as SOTIF \cite{iso21448}, the iterative data-based V-model refers to a complete coverage of the whole space by a sufficient decomposition through trigger constraints, yet in a data-based way. Moreover, the strategic generation of synthetic data offers the possibility of uncovering trigger conditions at an early stage and countering the challenge of "unknown unknowns". For example, logical inference based on existing trigger conditions can be used for this purpose.  

\textbf{System Development:}  This level represents the final implementation, which takes up and connects the previously decoupled paths of the architecture and data, and yields to a resulting system. Thereby, the system can consider individual dedicated (AI) models up to a system of systems, depending on the task. In addition, with regard to the overall process reference model, the system development phase completes the left leg of the V, which represents the design phase, and forms the intersection with the V\&V phase, the right leg of the V that is integral to the iterative data-based V-model.

\textbf{System Evaluation:} The first evaluation stage within the V\&V phase is represented by system evaluation. Here, the previously stipulated (functional) requirements are verified on the basis of the test split of the generated simulation data. Accordingly, this first evaluation stage can also be referred to as a virtual system testing. In more detail, this simulative performance evaluation is carried out using specified trigger conditions provided in terms of the dataset and assesses performance through corresponding indicators and acceptance criteria, which have to be derived from the individually customizable safety argumentation. 

\textbf{Real World Dataset Generation:} In this context, real-world data is generated. Data generation takes place in accordance with the data design domain. For safety reasons, the real world dataset represents a subset of the simulation dataset. Moreover, the effort in the real world can be limited for reasons of cost-effectiveness if this is permitted by the safety argument. As a consequence, the dedicated requirement for the scope of the real world dataset depends on the individual functionality to be developed and the corresponding safety argumentation and assurance. Therefore, a general specification cannot be provided by the process reference model.

\textbf{System Transfer:}  In order to counteract the gap between simulation and real world, the system transfer stage is introduced explicitly. The systems functionality can thus be adapted on the basis of the real data from the previous stage.

\textbf{Transfer Evaluation:} This second evaluation stage within the V\&V phase constitutes the transfer evaluation of the system. Here, the previously stipulated requirements are verified on the basis of the test split of the generated real world data. Consequently, this second evaluation stage can also be described as a simulation-based system test, which is carried out by means of real world data. The procedure as well as the performance indicators and acceptance criteria can be adopted from the first evaluation stage, the system evaluation, as the main difference is the changed origin of the data. Consequently, the desired functionality of the system and the associated quality and assessment criteria can be adopted.

\textbf{Open-Loop Evaluation (Silent Testing):}  This is the third evaluation stage, which marks the transition back to the real world. More precisely, the developed system is evaluated in the real world with the real input, whilst the system output is not applied in the real world. This open-loop evaluation thus represents a first step in the gradual return to the real world. In recent years, the open-loop evaluation has received increasing attention and can also be interpreted as silent testing \cite{wang2021online}, shadow mode, and virtual assessment of automation in field operation (VAAFO) \cite{wang2020reduction}.

\textbf{Closed-Loop Evaluation:} In the context of the fourth evaluation stage, the term closed-loop refers to the previous open-loop, which is now closed, meaning that the evaluation formalizes the system-based feedback into the real world. Here, the functionality is initially verified in the real world. Furthermore, this closed-loop evaluation phase is best viewed as a real world system test or, with regard to the application of automated driving, as a vehicle-in-the-loop (VIL). In the case of automated driving, this system evaluation can be carried out on the proving ground, for example. 

\textbf{Field Operation Evaluation:} The fifth and final evaluation phase is carried out in the real world on a larger scale than the previous one and aims to validate the system in the real world. It can be seen that V\&V, which was mostly considered jointly, is separate in the real world. While verification in the real world is possible on the proving ground, for example, validation requires further operation in the field. In principle, this kind of evaluation setting matches the consideration of on-the-road tests in automated driving applications.

\textbf{Deployment - System Operation \& Monitoring:} Once the system has successfully passed all evaluations, the acceptance criteria stipulated by the safety argumentation are fulfilled and the system can be depolyed. Continuous trust building arise in the course of system operation and ongoing monitoring. Furthermore, intelligent data harvesting can be performed during operation. In automated driving, for example, each individual vehicle can collect and select local data such that the cumulative gain can be used for subsequent refinement measures. 

\textbf{Detection of Deficiencies:} By means of the safety argumentation, corresponding acceptance criteria, and system specific performance indicators, deficiencies can be detected. If, e.g., an "unknown unknown" is detected at the overall level, which can have or has had catastrophic consequences, the approval of the systems must be withdrawn and the development and assurance process must cycled again from the beginning, taking into account the adapted requirements. While this is defined here in a structured way, we can see a very similar approach in practice today in the example of Cruise LCC in the USA, according to the incident of hitting a pedestrian \cite{equipmentrecallreport, NHTSARecall23E, NHTSARecallLetter}.

\textbf{Continuous Refinement:} In case of insufficencies, the overall loop is reinitiated by starting with a refinement of the ODD. A continuous transfer of the acquired findings back into the simulation and a step-by-step return to the real world is an integral part of the framework. Even in the long term, the use of synthetically generated data is targeted for safety, and efficiency reasons. In terms of trigger conditions, real world data might be limited to uncovered trigger conditions, while synthetic data allows the consideration of conceivable but unseen trigger conditions. Thus, the use of exclusively real data represents a subspace of imaginable trigger conditions and is therefore only effective in combination with synthetically augmented data. Accordingly, synthetic data can increase safety and efficiency in the development process. In general, moreover, the continuous refinement of the framework leverages the iterative nature of error analysis and specification adaptation, which increases safety and efficiency throughout the entire product lifecycle.

\textbf{Failure Handling:} Only the forward-looking transitions are indicated in Figure \ref{fig:ours}. Nevertheless, a large number of subordinate feedback flows are present, which have been omitted for the sake of clarity. Dealing with failed evaluations is of particular importance, which is why this is described in more general terms. 

On the one hand, an error can be caused by the design and realization of the functionality. On the other hand, it can be caused by gaps in the data used. Analyzing the error case can provide more information about the underlying cause. If an error case is caused by data and the data is within the data design domain, there is an error in the dataset generation. However, if the error case data is part of the data-specific ODD but not the data design domain, there is a gap in the specification of the data design domain. Otherwise, if the error case data is neither part of the data-specific ODD, the data-specific ODD definition itself must be adapted. An adaptation of a specification, e.g., the data-specific ODD or the data design domain, always requires stepping back to this level and revising the subsequent levels.

However, the functionality itself can also be the source of an error. If the selected function is within the architectural design domain, it is conceivable that the choice of architecture, although permissible, was not appropriate. If the realized functionality falls within the data-specific ODD but is not covered by the architectural design domain, the architectural design domain must be updated. If the function causes a behavior that does not correspond to the data-specific ODD but does correspond to the function-specific ODD, the data-specific ODD must be updated. Ultimately, gaps in the function-specific ODD can also lead to a subsequent error, which may require this specification to be updated. Furthermore, it is generally assumed that the assessments are conducted within the ODD. Otherwise, an ODD refinement is deemed to take place.

\textbf{Safety Argumentation Decoupling:} Along the process outlined above, the safety argumentation and safety assessment are customizable. While the VVM project \cite{VVMOverall} and Waymo's safety determination lifecycle \cite{favaro2023building} specify the absence of unreasonable risk in detail with respect to the system, this is omitted by the iterative data-based V-model, analogous to the classical V-model \cite{brohl1993v}. Nevertheless, it is evident that distinct evaluation stages throughout the V\&V process outlined above require the definition of acceptance criteria based on a system-specific safety argumentation. In this way, Waymo's case credibility assessment \cite{favaro2023building} is built-in by design, albeit with greater flexibility. This is due to the fact that the safety argumentation is not predefined. Therefore, to achieve the overall objectives, the framework enforces the creation, evaluation, and refinement of the safety argumentation. In particular, this customizability of the safety argumentation within the proposed framework implicitly leads to reasonableness, confidence, and coverage assessments along the interative refinement and the corresponding approval, similar to Waymo's case credibility assessment. Since the safety argumentation depends on the system as well as its environment and context, in other words specific on the ODD, and is also subject to application-specific standards and regulations, the safety argumentation is required to be decoupled and customizable to achive the desired generality of the methodology.

\subsection{Classification and Delimitation of the Methodology}\label{sec:classi}

The framework of the proposed iterative data-based V-model takes up existing further developments of the V-model \cite{VVMOverall} as well as innovative frameworks \cite{favaro2023building, karpathy_cvpr21} for handling complex systems. It addresses the characteristics of complex systems that integrate AI. This is illustrated by the use of data-based methods and the separate consideration of the architecture. In particular, complex systems are handled by means of dedicated stages and the formalized exploitation of simulation and real data. Central aspects from the AI Family project \cite{KIFamilie}, in particular the two sub-projects AI Delta Learning \cite{KIDeltaSynData} and AI Data Tooling \cite{KIDataTooling}, are thus taken up and formally integrated into an overall perspective. At the same time, the iterative data-based V-model offers the opportunity to take the results of the above-mentioned projects \cite{KIFamilie, KIDeltaSynData, KIDataTooling} into account. For instance, the stringent safety argumentation that was developed for a pedestrian detection AI \cite{KIAbsicherungSynMethoden, KIAbsicherungSynAbsicherung} can be considered. In addition, approaches for generating synthetic data \cite{KIAbsicherungSynData, KIDeltaSynData} can also be incorporated. This illustrates the generic and unifying character of the iterative data-based V-model framework. Moreover, the framework offers the possibility of using the emerging imaginative intelligence \cite{wang2024does, li2024sora} in the development process and is therefore also equipped for future developments.

Furthermore, the transition from a scenario-based approach to a general data-based approach opens up the broad applicability of the methodology. Additionally, the present approach enables scalability by the consideration of different levels of granularity of a system and, thus, also the system complexity. Thereby, the methodology addresses the architecture layer, the behavioral layer, as well as the in-service operational layer of Waymo \cite{karpathy_cvpr21} or the capability, engineering, and real world layer of the VVM project \cite{VVMAPerspectives}. Due to the claim of a generic framework and the necessity of individual performance measures and acceptance criteria, the framework does not claim to be a comprehensive framework for safety assessment and assurance. Nevertheless, Waymo's Case Credibility Assessment \cite{favaro2023building} approach is inherently integrated. This is due to the fact that the individual definition of the safety argumentation and thus the acceptance criteria and performance indicators require a suitability assessment, while the multiple assessments under increasingly realistic conditions require to address the coverage assessment. The iterative approach implies a process refinement within the structure defined by the framework. Hence, a native implementation of credibility \cite{koopman2019credible} can be envisioned. In particular, with increasing time and scale of the system, validation of the process and product takes place, thus addressing trust and credibility. This analysis of the proposed framework in relation to the improved V-model \cite{VVMOverall} of the VVM project, the safety determination lifecycle \cite{favaro2023building} of Waymo, and the data engine \cite{karpathy_cvpr21} of Tesla demonstrates that the iterative data-based V-model specifically combines the various methods and perspectives and formalizes them at the macro process level in order to maintain the generality of the classical V-model \cite{brohl1993v}.

While the framework does not explicitly address safety assessment and assurance, it opens up the possibility of data-driven functional safety assurance along with the potential to incorporate the results of current research from projects such as SUNRISE \cite{SUNRISE} and V4SAFETY \cite{V4SAFETY}. While allowing for statistical assurance of AI systems, the framework can also be applied to more traditional approaches. Ultimately, this leaves space for the use of different concepts. This is particularly important with regard to safety and explainability of AI systems, as research in this area is still raising open questions and different solutions are conceivable \cite{neto2022safety}. The future applicability of the process reference model is therefore addressed by the safety argumentation flexibility.

For an extended analysis of the iterative data-based V-model in relation to the previously analyzed process reference frameworks from Section \ref{02_New}, major characteristics are compared in Table \ref{tab:compare_frameworks}. Thereby, the advantages and disadvantages of the individual frameworks are illustrated in a compact and abstract manner while demonstrating that the iterative data-based V-model is able to unite the various frameworks except for the safety assessment, which is purposefully detached in accordance with the classical V-model. The individual criteria in Table \ref{tab:compare_frameworks} result from the analysis of the innovative development processes from Section \ref{02_New}, in particular from the discussion as well as the fundamental principles from Section \ref{03_Methodology}.

More specifically, the proposed framework extends the improved V-model specifically towards a continuous integration process that allows to respond to changes in the real world and address the open long-tail distribution challenge over time. Along with this standardization of the different frameworks, there is also a simplification. As discussed above, compared to VVM and Waymo, due to the chosen abstraction of the process, only one perspective is required to address different levels of system's granularity and complexity. Likewise, data-based processes, such as the one of Tesla's data engine, can be considered upfront, whilst a variety of system types, such as traditional or and mixed systems, can be considered in parallel. This is supplemented in Table \ref{tab:compare_frameworks}, which contains a more detailed assessment of the proposed framework in relation to previously investigated process reference frameworks. 

\newcommand\RotText[1]{\rotatebox{90}{\parbox{3.9cm}{\raggedright#1}}}

\begin{table}[]
	\centering
	\caption{Comparison of the previously analyzed process reference frameworks with the established iterative data-based V-model.}
	\begin{tabularx}{\linewidth}{l *{5}{>{\raggedright\arraybackslash}X}}
		\toprule
		& \RotText{Classical V-model} & \RotText{Improved V-model \quad \quad \quad \quad \quad  \tiny(VVM Project)} & \RotText{Safety Determination Lifecycle \tiny(Waymo)} & \RotText{Data Engine \quad \quad \quad \quad \quad \quad \quad \quad } & \RotText{Iterative data-based V-model \tiny(proposed)} \\
		\midrule
		Design phase & \cmark & \cmark & (\cmark) & \cmark & \cmark \\
		V\&V phase & \cmark & \cmark & \cmark & \cmark & \cmark \\
		Application independence & \cmark & \xmark & (\cmark) & (\cmark) & \cmark \\
		\midrule
		Generic across system granularities & \cmark & \xmark & \xmark & (\xmark) & \cmark \\
		Use of specific databases & \xmark & \cmark & (\cmark) & \cmark & \cmark \\
		Use of generic databases & \xmark & \xmark & (\xmark) & \xmark & \cmark \\
		Formalized simulation exploitation & \xmark & \xmark & \xmark & \xmark & \cmark \\
		\midrule
		Iterative product refinement & \xmark & \xmark & \cmark & \cmark & \cmark \\
		Iterative process refinement & \xmark & \xmark & \cmark & (\xmark) & (\cmark) \\
		Continuous trust building & \xmark & \xmark & \cmark & (\xmark) & \cmark \\
		\midrule
		System monitoring & \xmark & \xmark & \cmark & \cmark & \cmark \\
		Safety assessment (e.g. w.r.t. residual risk) & \xmark & \cmark & \cmark & \xmark & \xmark \\
		Safety argumentation customizability & \cmark & \xmark & \xmark & (\xmark) & \cmark \\
		\midrule
		Appropriate for traditional systems & \cmark & \cmark & \cmark & \xmark & \cmark \\
		Appropriate for AI systems & \xmark & (\cmark) & (\cmark) & \cmark & \cmark \\
		Appropriate for complex systems incl. AI & \xmark & (\cmark) & \cmark & (\cmark) & \cmark \\
		\midrule
		Overall generality and transferability & \cmark & (\xmark) & (\cmark) & (\cmark) & \cmark \\
		Overall suitability for emerging AI systems & \xmark & (\cmark) & (\cmark) & (\xmark) & \cmark \\
		\bottomrule
	\end{tabularx}%
	\label{tab:compare_frameworks}
\end{table}%

\subsection{Summary and Discussion of the Proposed Framework}

Consequently, it can be summarized that the iterative data-based V-model
\begin{itemize}
	\item represents a systematic update of the classical V-model,
	\item recognises the data-driven AI development and V\&V,
	\item generalizes across innovative development frameworks,
	\item thus formalizes a unifying process reference model.
\end{itemize}

Therefore, like the classical V-model, the iterative data-based V-model
\begin{itemize}
	\item decouples the process from specific safety assurance,
	\item unites multiple perspectives and approaches into a single,
	\item applies across different levels of system granularities,
	\item thus enables the desired generality and transferability.
\end{itemize}

In addition, the various advantages of the different innovative development processes are taken into account by the iterative data-based V-model, that 
\begin{itemize}
	\item considers system environment/context through the ODD,
	\item maps the open world in a managable (trigger) datasets,
	\item accounts for the prospective and retrospective view,
	\item enables iterative refinement throughout the lifecycle,
	\item thus harmonizes advantages of existing methodologies.
\end{itemize}

Furthermore, the data-based iterative V-model also addresses respective disadvantages of existing process reference models, as it
\begin{itemize}
	\item relaxes assumptions w.r.t. databases or hardware,
	\item formalizes the exploitation of simulation and real world,
	\item applies to traditional, AI-based, or mixed systems,
	\item represents an application-agnostic methodology,
	\item and overcomes limitations of existing frameworks.
\end{itemize}

The iterative data-based V-model harmonizes the respective advantages from Table \ref{tab:compare_frameworks_ad_disad} while counteracting the disadvantages from Table \ref{tab:compare_frameworks_ad_disad}. Thereby, the classical V-model is deliberately extended to the needs of complex systems incorporating AI. 

The proposed framework also has some implications. Particularly, while decoupling the framework from safety assessment and argumentation enables broad applicability, it does not inherently ensure the development of safe systems. Therefore, for safety-critical applications, individual safety assessments and arguments are necessary. However, the framework provides the desired flexibility and adaptability. Morover, among other things, the harmonization of different approaches has a specific implication on the formalization of the process, which is illustrated in Figure \ref{fig:ours}. This formalization explicitly addresses various data sources, from synthetic to real data. In particular, the systematic consideration of synthetic data facilitates the early uncovering of trigger conditions and counteracts the challenge of “unknown unknowns”. The framework thus enables greater efficiency and safety through the systematic consideration of synthetic data. Furthermore, the continuous refinement provided by the framework and thus the iterative nature of error analysis and specification adaptation throughout the entire product lifecycle ensures and increases safety.

Beyond that, the formalization also entails the separation in the depth of the design domain and design implementation as well as in the separation in the breadth of the architecture and the parameterizing data, in order to systematically close gaps in development phase. Overall, a unique feature of the framework is its ability to cover different types of systems, levels of granularity and complexity, and application areas from a unified perspective. The property to adapt to varying system levels and complexities is illustrated abstractly in the following.

%% file: src/04_Examples.tex
\section{Sketch of Possible Applicatons of the Iterative Data-based V-model}\label{04_Examples}

In this section, the application of the proposed iterative data-based V-model is illustrated by means of automated driving at various levels of granularity, thus also the system's complexity. Thereby, the application illustration does not represent practical examples. Rather, the illustration serves as an academic explanation of the usability and possible practical application of the introduced process reference framework. Consequently, the focus is on the description of the process, i.e., how to proceed according to the process reference framework, and not on the result of a specific example. Moreover, a comprehensive description of a practical application would entail a detailed description of the system requirements, the database, etc., which would go way beyond the scope of this paper and impair clarity. As a result, this section concentrates on the application illustration of the process.

In this context, the generally described functionality and the usability across different system levels is sketched. The following Table \ref{tab:exampl_design} illustrates the design phase, while continued Table \ref{tab:exampl_VV} focuses on the V\&V phase of the iterative data-based V-model. These examples outlined in the Table \ref{tab:exampl_design} and continued part of Table \ref{tab:exampl_VV} provide high-level considerations and do not claim to be exhaustive. Rather, the purpose of these illustrations is to provide a clear and understandable intuition for the application of the methodology outlined in Section \ref{03_Methodology}. The application of automated driving within a highway operation ODD is used for this purpose. Subsequently, the levels considered comprise the overall system, here the automated driving (AD) stack, a dedicated subsystem of the overal system, namely the perception and the lidar detector component as one of the sensor components for automated driving.

On the one hand, this demonstrates how the framework can be applied at different levels of granularity. On the other hand, it illustrates that Waymo's layered approach \cite{webb2020waymo} and the multi-perspective approach \cite{VVMAPerspectives} of the VVM project are implicitly taken into account in an application-independent single-perspective methodology. This is reflected, for instance, in the closed-loop evaluation at the system and component level, which enables the identification of necessary improvements at the overall system level. Apart from this particular case, it can also be seen in general how, in addition to refinement within a system level, refinement is organized across different system levels.

\begin{table*}[p!]
	\centering
	\caption{Potential application of the Iterative Data-based V-model to Automated Driving Applications at different Levels of Granularity.}
	\begin{tabularx}{\linewidth}{p{3.6cm} *{3}{>{\raggedright\arraybackslash}X}}
		\toprule
		\diagbox[width=4cm, height=1.5cm]{\textbf{Stages}\\\textbf{of the iterative} \\ \textbf{data-based V-model}}{\textbf{Automated driving at}\\\textbf{different levels of}\\\textbf{granularity}} & \multicolumn{1}{c}{\textbf{AD Stack}} & \multicolumn{1}{c}{\textbf{Perception}}  & \multicolumn{1}{c}{\textbf{Lidar Detector}} \\
		\toprule
		\textbf{Operational Design Domain (ODD)} &
		\multicolumn{3}{>{\hsize=\dimexpr3\hsize+4\tabcolsep+\arrayrulewidth\relax}X}{ $\bullet$ German highways; $\bullet$ ego vehicle speed of up to 120 km/h; $\bullet$ other road users speed of up to 160 km/h; $\bullet$ from traffic jams to high-speed traffic; $\bullet$ varying traffic densities and weather conditions;}\\
		\midrule
		\textbf{Function-specific ODD} 
		& $\bullet$ requirement specification of sensing, planning and acting functionalities within the ODD; $\bullet$ maneuvers (overtaking, lane changing, merging, exiting) within the ODD and legal requirements; $\bullet$ different conditions (varying weather, traffic jams, emergency vehicles, pedestrians along the highway ODD);  
		& $\bullet$ object detection, classification and tracking within the ODD; $\bullet$ obstacle and lost cargo detection; $\bullet$ lane and free space detection; $\bullet$ detection and classification of traffic signs and traffic lights;
		& $\bullet$ object and free space detection within the ODD; \\
		\midrule
		\textbf{Data-specific ODD} 
		& $\bullet$ requires perception data with a range of at least 200 meters to the front and rear and 100 meters to the left and right; $\bullet$ vehicle state estimates (position in global coordinates, velocities, accelerations in the navigation coordinates, heading and yaw rate relative to the navigation coordinate system); $\bullet$ highway HD maps; $\bullet$ information processing at a rate of 10 Hz; 
		& $\bullet$ requires vehicle state estimation in standard signal rates and and characteristic signal noise; $\bullet$ provide an environment model for the region of interest in order to make informed behavioral decisions; $\bullet$ information processing at a rate of 10 Hz; 
		& $\bullet$ requires lidar raw data with characteristic signal noise and a range coverage of 150 meters; $\bullet$ provide object and free space detections at a rate of 10 Hz; \\
		\midrule
		\textbf{Architectural Design Domain} 
		&$\bullet$ domain of possible AD stack architectures (classic architectures with dedicated interfaces between individual submodules, architectures with orchestrators that take control of the interface, architectures that are fully end-to-end learned); 
		& $\bullet$ domain of possible perception architectures (permanent 360-degree perception, situation dependent perception); 
		& $\bullet$ domain of possible lidar detector architectures (range of deep neural network architectures); \\
		\midrule
		\textbf{Architecture Definition} 
		& $\bullet$ definition includes a specific instance from the range of possible architectures (e.g. an innovative architecture based on an orchestrator); $\bullet$ deployment plan on individual ECUs; 
		& $\bullet$ definition represents a specific case from the spectrum of possible architectures (e.g. situation dependent perception); $\bullet$ consideration of the use on individual ECUs; 
		& $\bullet$ definition determines the choice of  the deep neural network architecture within the given spectrum; $\bullet$ includes architecture-specific definitions (layers, activation functions, properties of the head, e.g., deterministic or probabilistic outputs);  \\
		\midrule
		\textbf{Data Design Domain} 
		& $\bullet$ defines general requirements for the creation of datasets both in simulation and the real world; $\bullet$ requires the generation of data according to the data-specific ODD; $\bullet$ covering different velocities, traffic densities, various times of day, and changing environmental conditions; $\bullet$ data from traffic jams, emergency situations, pedestrians within the ODD, corner cases (e.g. accidents, animals on the road); $\bullet$ defines train test split;
		& $\bullet$ demands the generation of sensor and state estimation data according to the data-specific ODD; $\bullet$ coverage of different velocities, traffic densities, different times of day, changing environmental conditions; $\bullet$ data from traffic jams, emergency situations, pedestrians within the ODD, corner cases; $\bullet$ defines train test split;
		& $\bullet$ specifies the sensor type, the field of view, the amount of layers and the mounting position to fulfill the data-specific ODD; $\bullet$ specifies lidar resolution, frame rate, and latency; $\bullet$ coverage of various environmental conditions (weather, traffic) within the ODD; $\bullet$ defines train test split;\\
		\midrule
		\textbf{Sim. Dataset Generation} 
		& $\bullet$ involves systematically generating simulation data considering, sensor specifics, vehicle dynamics, traffic simulation, scenario generation; $\bullet$ generates data with specific resolution, frame rate, and latency; $\bullet$ required to ensure the scenario coverage matches the requirements of the data design domain;
		& $\bullet$ generates simulation data by means of a simulation software considering, realistic objects, obstacles, lanes, free spaces and sensor models; $\bullet$ generates data with ground truth annotations, defined resolution, frame rate, and latency; $\bullet$ covers cases according to the requirements specification of the data design domain;
		& $\bullet$ generates data according to the data design domain with a specific scan frequency; $\bullet$ consideres physics of lidar sensors, including laser beam emission, reflection, and reception, to generate accurate point cloud representations; $\bullet$ accounts for sensor noise and environmental influences (fog, rain and blending);
		\\
		\midrule
		\textbf{System Development} 
		& $\bullet$ realization of the automated driving stack including various submodules with respect to the architecture definition; $\bullet$ use of the created simulation scenarios and corresponding data obtained from the sim. dataset generation;
		& $\bullet$ implementation and realization of the selected perception architecture; $\bullet$ use of the created simulation and corresponding data derived from the sim. dataset generation;
		& $\bullet$ parameterization and implementation of a selected lidar detector neural network according to the architecture definition; $\bullet$ training of the neural network using the data from the sim. dataset generation; \\
		\bottomrule	
	\end{tabularx}
	\label{tab:exampl_design}
\end{table*}

\begin{table*}[h!]
	\ContinuedFloat
	\centering
	\caption{Potential application of the Iterative Data-based V-model to Automated Driving Applications at different Levels of Granularity (continued).}
	\begin{tabularx}{\linewidth}{p{3.6cm} *{3}{>{\raggedright\arraybackslash}X}}
		\toprule
		\diagbox[width=4cm, height=1.5cm]{\textbf{Stages}\\\textbf{of the iterative} \\ \textbf{data-based V-model}}{\textbf{Automated driving at}\\\textbf{different levels of}\\\textbf{granularity}}&	\multicolumn{1}{c}{\textbf{AD Stack}} & \multicolumn{1}{c}{\textbf{Perception}}  & \multicolumn{1}{c}{\textbf{Lidar Detector}} \\
		\toprule
		\multirow{2}{*}{\textbf{System Evaluation}}
		& $\bullet$ performs evaluation on the test dataset split defined within the data design domain definition; $\bullet$ considers among others, functional testing of the realized automated driving stack, scenario testing, regulatrory compliance testing, etc.; 
		& $\bullet$ evaluates realized perception on the test dataset; $\bullet$ analysis of object detection accuracy on divers objects under various conditions; 
		& $\bullet$ evaluates object and free space detections under various ODD-specific circumstances; \\
		\cline{2-4}
		&\multicolumn{3}{>{\hsize=\dimexpr3\hsize+4\tabcolsep+\arrayrulewidth\relax}X}{
			$\bullet$ entails the simulative performance assessment based on performance indicators and acceptance criteria derived from the individual safety argumentation; $\bullet$ evaluates trigger conditions according to the specification of the data design domain and correspondingly generated data;
		} \\
		\midrule
		\textbf{Real World Dataset Generation} 
		& $\bullet$ gathering of relevant real world measurement data, vehicle dynamics characteristics, system latencies and actuator properties;
		& $\bullet$ collecting real raw sensor data required to create the environment model along specified real world state estimates;
		& $\bullet$ acquisition of real world data via a dedicated sensor module that meets the requirements; \\
		\cline{2-4}
			&\multicolumn{3}{>{\hsize=\dimexpr3\hsize+4\tabcolsep+\arrayrulewidth\relax}X}{ $\bullet$ generates data from the real world corresponding to the respective data design domain; $\bullet$ coverage of the data design domain is defined system dependant according to the individual safety argumentation that considers risks and costs of generating real world data, particularly with regard to corner cases and hazardous situations;
		} \\
		\midrule
		\textbf{System Transfer} 
		& $\bullet$ adjusting individual (sub-)modules of the automated driving stack and their interaction by means of acquired real-world data to compensate the gap between simulation and real world;
		& $\bullet$ specifics within the chosen perception architecture are adjusted using generated real world data; 
		& $\bullet$ parameterization of the lidar detector's neural network is adapted, e.g. by fine-tuning using real data;  \\
		\midrule
		\textbf{Transfer Evaluation}	
		& $\bullet$ performance assessment evaluation of adapted automated driving stack system using real data in alignment with the individual safety argumentation as in the previous evaluation; 
		& $\bullet$ evaluation of the performance of object detection, classification, tracking, obstacle detection, lost cargo detection etc. using the respective real world test dataset split; & $\bullet$ evaluation of the performance of the lidar detector based on the test split of the generated real data; \\
		\midrule
		\textbf{Open-Loop Evaluation (Silent Testing)} 
		& $\bullet$ evaluation of sensing, planning, acting and interactions of functionalities on real hardware in the real world in partial separation to prevent deployment in the real world while still being able to evaluate the capabilities of the AD stack;
		& $\bullet$ evaluation of the perception systems object detection, classification, tracking etc. in the real world without feeding the output back into the overall system;
		& $\bullet$ evaluation of the lidar detector in the real world without further use of the output in other functionalities; \\
		\midrule
		\textbf{Closed-Loop Evaluation} 
		& $\bullet$ evaluation of sensing, planning, acting functionalities on real hardware in the real world in full functional interplay and sytem's interaction with the real world; $\bullet$ limited to operations within the proving ground; 
		&\multicolumn{2}{>{\hsize=\dimexpr2\hsize+3\tabcolsep+\arrayrulewidth\relax}X}{
			$\bullet$ evaluation is limited to operation within the proving ground; $\bullet$ perception and lidar detection, unlike planning, is usually open-loop and therefore should not lead to changes; $\bullet$ if this closed-loop evaluation differs from the open-loop evaluation from the previous step, it indicates that related components within the automated vehicle stack need to be refined;
		} \\
		\midrule
		\textbf{Field Operation Evaluation} &\multicolumn{3}{>{\hsize=\dimexpr3\hsize+4\tabcolsep+\arrayrulewidth\relax}X}{ $\bullet$ previous restriction to the proving ground is removed; $\bullet$ performance assessment takes place in field operation within the ODD; $\bullet$ enables a broader coverage of possible test scenarios and influences; 
		} \\
		\midrule
		\textbf{Deployment - System Operation \& Monitoring} &	\multicolumn{3}{>{\hsize=\dimexpr3\hsize+4\tabcolsep+\arrayrulewidth\relax}X}{ $\bullet$ with increasing time and scale of use, continuous trust builds up in the system; $\bullet$ continuous observation and analysis of system performance; $\bullet$ conducting intelligent data harvesting during operation; 
		} \\
		\midrule
		\textbf{Detection of Deficiencies} &	\multicolumn{3}{>{\hsize=\dimexpr3\hsize+4\tabcolsep+\arrayrulewidth\relax}X}{ $\bullet$ detection of deficiencies on the basis of a strong link between system observation and safety argument \& acceptance criteria based identification; $\bullet$ uncovering of "unknown unknowns"; $\bullet$ withdrawal of approval, if necessary; $\bullet$ design and V\&V cycle reinitiation 
		} \\
		\midrule
		\textbf{Continious Refinement} &	\multicolumn{3}{>{\hsize=\dimexpr3\hsize+4\tabcolsep+\arrayrulewidth\relax}X}{ $\bullet$ deficiencies uncovered during operation are fed back into the process and the product via a new overall cycle; $\bullet$ findings are transferred to the simulation, the product is refined and gradually reintroduced into the real world; $\bullet$ exploits the safety and efficiency-related benefits of simulation; $\bullet$ ensures uniform safety standards in every development and improvement cycle;
		} \\
		\bottomrule	\end{tabularx}
	\label{tab:exampl_VV}
\end{table*}

%% file: src/05_Conclusion.tex
\section{Conclusion}\label{05_Conclusion}

The iterative data-based V-model is a process reference model with the generic claim of the classical V-model. However, the challenges of complex systems that include AI are explicitly addressed in order to provide a development and V\&V framework that can serve as a generic guideline. In this way, the iterative data-based V-model can be adapted to current and future challenges.

On the one hand, the development process of AI systems is mapped in a more explicit form, without excluding traditional systems. On the other hand, greater emphasis is placed on data, which is firmly integrated into the process. The data makes it possible to bring the product and the process, the real world and the simulation world, as well as the prospective and the retrospective perspective to a common ground. This enables the diverse perspectives of other frameworks to be mapped into one application-independent single-perspective framework. In addition, the iterative data-based V-model allows different granularities of the system level to be taken into account and thus a stronger generality. Although cognitive cyber-physical systems such as autonomous systems (e.g. fully automated vehicles) are explicitly addressed, simpler and less safety-critical systems can also be developed and validated by means of this framework.

The iterative data-based V-model is also characterized by the fact that statistical methods can, but do not have to, be given greater consideration during development, verification, validation, and continuous improvement. Openness to traditional safety argumentation methods and to statistical methods that take account of the inherent complexity is equally guaranteed.

In addition, the challenges of the open world are taken into account through the iterative process, thus incorporating the continuous integration approach. Furthermore, this represents a V\&V of the process and the associated customizable safety argumentation and ultimately empowers a comprehensive safety assurance. In particular, the retrospective approach makes it possible to strengthen confidence in the completeness of the data, the performance metrics, and the release criteria, in other words in the product and the process, and ultimately in the ambitious goals that have been defined.

To summarize, the iterative data-based V-model takes up a variety of different approaches and methods and formalizes the fundamental ideas. This requires a decoupling from the safety argumentation and at the same time opens up the consideration of various applications and procedures, similar to the classical V-model.